\documentclass[preprint2]{aastex}

\newcommand {\nh}{N$_{2}$H$^{+}$\ }
\newcommand {\nhns}{N$_{2}$H$^{+}$}	
\newcommand {\co}{C$^{18}$O\ }
\newcommand {\cons}{C$^{18}$O}		
\defcitealias{Kirk06}{Paper~I}
\newcommand{\hk}{}

\begin{document}
\title{Dynamics of Dense Cores in the Perseus Molecular Cloud}


\author{Helen Kirk\altaffilmark{1,2}, Doug Johnstone\altaffilmark{1,2}, and Mario 
	Tafalla\altaffilmark{3}}
\altaffiltext{1}{Department of Physics \& Astronomy, University of Victoria, 
	Victoria, BC, V8P 1A1, Canada; hkirk@uvastro.phys.uvic.ca}
\altaffiltext{2}{National Research Council of Canada, Herzberg Institute of 
	Astrophysics, 5071 West Saanich Road, Victoria, BC, V9E 2E7, 
	Canada}
\altaffiltext{3}{Observatorio Astron\'omico Nacional (IGN), Alfonso XII 3,
	E-28014 Madrid, Spain}

\begin{abstract}
We survey the kinematics of over one hundred and fifty candidate (and 
potentially star-forming) dense cores in the Perseus molecular cloud 
with pointed \nhns(1-0)
and simultaneous \cons(2-1) observations.  Our detection rate of
\nh is 62\%, rising to 84\% for JCMT SCUBA-selected targets. 
In agreement with previous observations, we find that
the dense \nh targets tend to display nearly thermal linewidths, particularly
those which appear to be starless (using Spitzer data), indicating
turbulent support on the small scales of molecular clouds is minimal.
For those \nh targets which have an associated SCUBA dense core,
we find their internal motions are more than sufficient to provide support 
against the gravitational force on the cores.  
Comparison of the \nh integrated intensity
and SCUBA flux reveals fractional \nh abundances between $10^{-10}$ and
$10^{-9}$.
We demonstrate
that the relative motion of the dense \nh gas and the surrounding
\co gas is less than the sound speed 
in the vast majority of cases ($\sim$90\%).  The point-to-point
motions we observe within larger extinction regions appear to be
insufficient to provide support against gravity, although we sparsely
sample these regions.  
\end{abstract}

\section{INTRODUCTION}
Stars form in the densest regions of a hierarchy of structures that exist 
within a molecular cloud.  Supersonic motions dominate on the largest 
scales \citep[e.g.,][]{Larson81} but appear to be much reduced on the 
smallest scale, that of a preprotostellar core \citep[e.g.,][]{Benson89}.  
Understanding what physical processes are at play at each scale ranging
from the largest to the smallest is a challenge that both observers
and theorists face.  The challenge is complicated
by the fact that each molecular cloud possesses different properties --
local environment has a strong effect on the star formation process.
Some molecular clouds, such as Taurus, display isolated star-forming 
cores which are quiescent and have a low star formation efficiency,
while others such as Orion display clustered
star-forming cores and have more turbulent motions and a higher
star formation efficiency \citep[e.g.,][]{Cohen79}.  While clustering increases
the complexity of a region and hence the difficulty in understanding
what processes are at play, the majority of stars appear to form
in environments which are clustered to some degree, and hence it
is important to study such regions to understand the impact of clustering.
In this paper, we examine the Perseus molecular cloud -- a cloud which
is less clustered and confused than structure in the Orion molecular cloud, 
but is not as isolated and quiescent as the Taurus 
molecular cloud.  The Perseus molecular cloud consists of a chain of 
distinct, well-known small clustered environments 
in which stars are forming -- e.g., NGC1333 and IC348 \citep{Bally_handbook}.  
There is a wealth of data uniformly spanning this cloud, courtesy of
the Spitzer `c2d' Survey \citep{Evans03} and the COMPLETE (Co-Ordinated
Molecular Probe Line, Extinction, and Thermal Emission) Survey
\citep[see][for a summary of the publicly available data]{Ridge06}.
The continuum data from the latter survey (SCUBA thermal dust emission
and 2MASS near-IR extinction) allow for the determination of
the (column) density structure of the cloud.  In particular, 
we \citep[][hereafter Paper I]{Kirk06} characterized the environments 
in which dense cores (which could eventually evolve to form a star) 
were themselves able to form.  In doing so, we determined a set of 
constraints on
the column density structure of the cloud which simulations should
match.  The goal of the present work is to extend that set of density
constraints to a set of dynamical constraints.

To sample the behaviour of the dense gas in cores across the molecular
cloud, we performed pointed \nh observations on a set of locations
including candidate dense cores from submillimetre data, points of 
high visual extinction from the Palomar plates, and peaks of large-scale
extinction from 2MASS data.  With the addition of the Spitzer 
data \citep{Jorgensen07},
we are able to differentiate between unevolved dense starless cores
and their more evolved protostellar bretheren.  We couple {\hk these} data
with the existing continuum data discussed above in order to determine
the dynamical behaviour of the dense gas within the cloud on a variety
of scales.  Our data {\hk do} not provide the resolution or full sampling
of some other recent surveys \citep[e.g.,][]{Walsh06}, but {\hk have}
the advantage of providing (sparse) sampling across a much larger
area, and {\hk are} thus complimentary to these other works.

\section{SOURCE CATALOG}

\subsection{SCUBA submillimetre}
The Perseus molecular cloud has been mapped over roughly 
$\sim 3.5$ square degrees in the submillimetre (sensitive to thermal 
radiation by dust grains)
at 850~$\mu$m with the Submillimetre Common User Bolometer Array (SCUBA) at
the James Clerk Maxwell Telescope in Hawaii. 
In \citetalias{Kirk06}, we identified approximately fifty cores
in the SCUBA data using a 6\arcsec\
sampled map of all existing data in the cloud.  The majority of {\hk these} 
data {\hk were} archival \citep[see][]{Hatchell05}, but also included some `fast scan'
maps (with insufficient integration time to allow for the typical 3\arcsec\ 
sampling) taken as part of the COMPLETE project \citep{Ridge06}. 
\citetalias{Kirk06} compared the properties of the SCUBA cores 
to the environment
they inhabit through comparison to near IR extinction data with 2.5\arcmin\ 
resolution (see discussion of {\hk these} data below), hence the consistent
areal coverage was of greater importance than resolution.  All of
the submillimetre cores were
found in regions of previously identified star-formation (e.g., B1, NGC1333), 
and none were found in the fast-scan mapped regions.  For our current 
project, we utilize maps created with a finer 3\arcsec\ sampling which
allows for better separation of close cores and determination of core
properties such as radius.  Appendix A discusses the re-reduction of
the data we performed, and presents a full source catalog.  We identify
seventy two submillimetre cores above our usual S/N level 
\citep[these cores are also used in][]{Jorgensen07}, as well as 15
additional potential submillimetre cores which do not satisfy our
S/N criteria.

The original 6\arcsec\ sampled map formed the basis of our target list of
dense cores to observe in \nhns(1-0).  We supplemented this with a
list of potential SCUBA cores which fell below the detection
threshold of our core-identification procedure in order to have as 
complete a list as possible.  This resulted in 89 dense core
targets.  These target positions are listed in Table~\ref{targets}
along with the associated better-defined SCUBA core from this paper.

\subsubsection{Spitzer \& young protostars}
The Perseus molecular cloud was also one of the clouds surveyed by the
Spitzer `c2d' project \citep{Evans03}.  While the submillimetre
wavelengths covered by SCUBA are sensitive to the dusty envelopes
of both starless cores and enshrouded protostars, the IR wavelengths 
probed by Spitzer, particularly the shorter wavelengths, are sensitive
to the central accreting object and hence young protostars.  Thus Spitzer
observations are ideal in distinguishing between starless cores and
their more evolved bretheren. 
The full Spitzer dataset is described in \citet{Jorgensen06} and
\citet{Rebull07}, while a 
catalogue identifying young protostellar candidates using a combination of
Spitzer and SCUBA data is presented in \citet{Jorgensen07}.  Note that the
work of \citet{Jorgensen07} utilizes the 3\arcsec\ sampled SCUBA core list
included in Appendix~A.
Table~\ref{targets} also shows whether the targets are associated with
a young protostar as detected with Spitzer.

A second comprehensive catalog listing the protostellar and 
starless cores in Perseus has also recently been published by 
\citet{Hatchell07} where the sources were classified using SEDs fit to
IRAS, Spitzer, SCUBA, and Bolocam observations.  A significant
difference between the two catalogs is that the Hatchell catalog
used only the shorter wavelength Spitzer IRAC data, while the 
J\o rgensen catalog relied primarily on the longer wavelength
Spitzer MIPS data.
The bulk of the two 
protostellar catalogs agree, but the J\o rgensen catalog 
contains five protostars not in the Hatchell catalog
and the Hatchell
catalog contains seventeen protostars not in the J\o rgensen
catalog.  

Of the five sources only in the J\o rgensen catalog,
one was outside the region included in the Hatchell catalog,
two were classified as starless cores in the Hatchell catalog, and 
the remaining two had submillimetre emission below the
threshold for identification.  The J\o rgensen catalog had
three criteria for classification as a protostar -- a set of
colour criteria for a source across the IRAC (3.6, 4.5, 5.8~$\mu$m) 
and MIPS (24~$\mu$m) bands, spatial
coincidence of a MIPS 24~$\mu$m source with a SCUBA core, or detection
of a SCUBA core with a high central concentration (see Appendix~A
for the definition of concentration).  None of the
seventeen protostars only identified
in the Hatchell catalog are associated with a high submillimetre
central concentration, therefore inclusion in the J\o rgensen catalog
would require a detection in the MIPS 24~$\mu$m band plus fulfilment
of either
the colour criteria or SCUBA core association.  We searched the Spitzer data 
\citep{Rebull07,Jorgensen06,Evans03} around each of the
seventeen protostars only in the Hatchell catalog.  Twelve of the 
seventeen did not have MIPS detections with a signal to noise above
five (seven of these had close SCUBA cores while five did not).
Of the remaining five, none had close SCUBA cores or colours satisfying
the colour criteria of the J\o rgensen catalog.
Of the sources only in the J\o rgensen catalog, we observed four of 
the five and all had
detections (our sources \#21, 22, 74, and 95).  Of the sources only 
in the Hatchell catalog, we observed ten of the seventeen and had detections
for seven (our sources \#92, 94, 101, 109, 110, 121, and 150). 

A range of linewidths, centroid velocities, and peak
intensities were found for these sources, therefore any potential
mis-classification is unlikely to bias our results. 

\subsection{Extinction - 2MASS}
From the COMPLETE Survey, we also have a map of the total column density 
created from extinction
measures derived using the NICER technique \citep{Lombardi01}
on the 2MASS dataset \citep{Ridge06,Alves}.  
This technique utilizes near infrared reddening of background stars in 
three wavelength bands (two colour indices) in order to determine 
the total column density of dust.  The
resolution of this map is 2.5\arcmin\ and spans the range of A$_{V}= $3 - 11 
within the region mapped by SCUBA (Figures 2 and 3 of \citetalias{Kirk06}).
\citetalias{Kirk06} analyzed {\hk these} extinction data and identified structures 
in the column  density on two scales.  We 
utilized the maxima in the smaller-scale structures (the `extinction cores' 
of \citetalias{Kirk06}) to
identify a further 24 targets for our \nh survey.  In later sections
of this paper, we utilize the larger-scale structures (`extinction supercores'
of \citetalias{Kirk06}) to
define the environments in which the dense cores inhabit.  The extinction 
target positions,
along with the associated information from the extinction and submillimetre
maps are given in Table~\ref{targets}.

\subsection{Palomar Plates}
The final extension to the dense core candidate target list was taken from
visually-selected targets in the red POSS-II Palomar plates.  We focussed
on plates in regions devoid of SCUBA cores in order to maximize the
extent of our coverage of environments within the cloud.  This provided
an additional 44 targets.  The information on these targets is given
in Table~\ref{targets}.

\begin{figure}[p]
\includegraphics[height=11.5cm,angle=90]{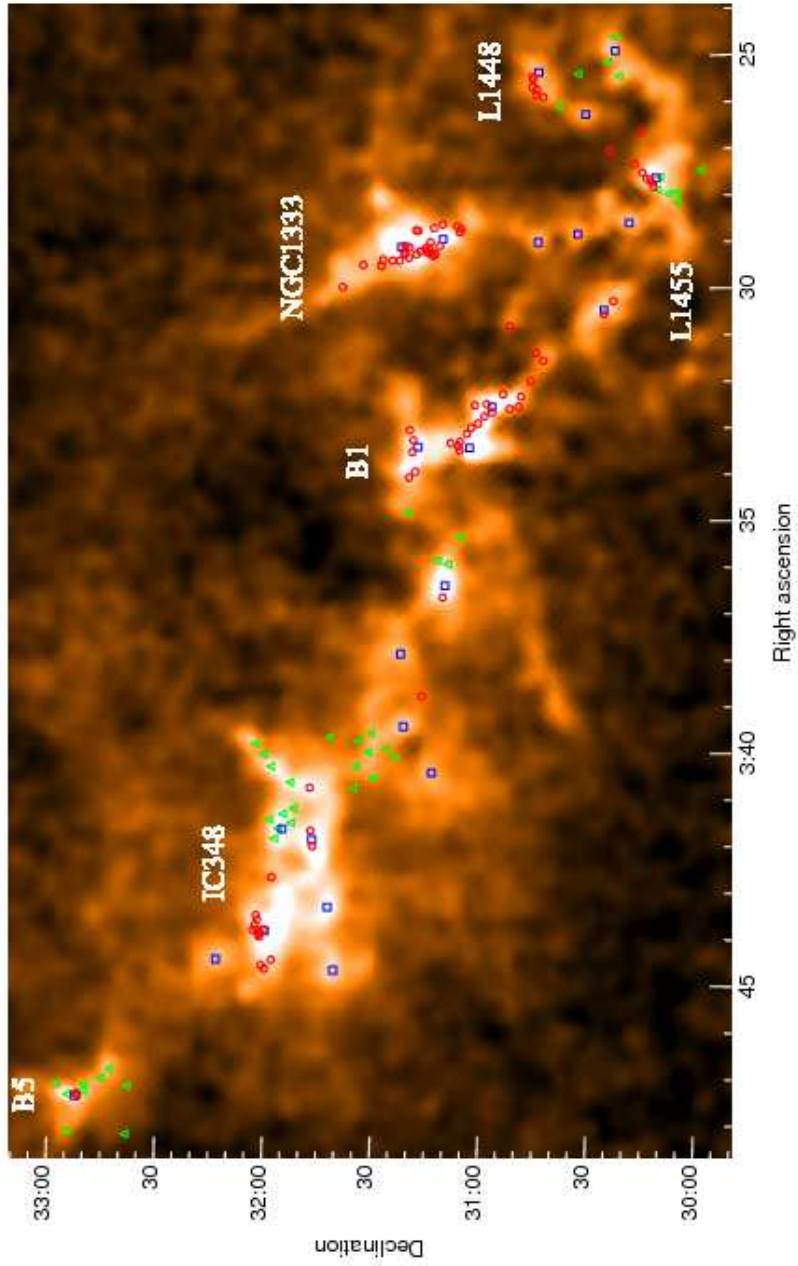}
  \caption{2MASS-derived extinction map of the Perseus molecular cloud 
	overlaid with the positions of our IRAM survey targets.  Red
	circles indicate the SCUBA-selected targets, green triangles
	indicate the Palomar plate-selected targets, and blue squares
	indicate the 2MASS-selected targets.  Well-known star-formation
	regions are labelled.
        }

  \label{fig_targets}
\end{figure}

Figure \ref{fig_targets} shows all of the targets we selected overlaid on the
extinction map of the Perseus molecular cloud.

\section{OBSERVATIONS AND DATA REDUCTION}
We made pointed observations of the 157 targets
using the 30~m IRAM telescope in Pico Veleta, Spain.  We observed
\nhns(1-0) in both polarizations using the (AB) 100~GHz receivers and
simultaneously observed \cons(2-1) in both polarizations using the (CD) 230~GHz
receivers, in both cases utilizing frequency switching.  We used the 
VESPA correlator and smoothed to 0.05~km/s channels.  The beamsize is
$\sim$25\arcsec\ for \nh and $\sim$11\arcsec\ for \cons.
Each observation was made for 2 minutes and had an rms of T$_{A}^{*}\sim$0.1~K.
We made multiple pointings on some of the cores to better resolve spectral
features or search for a signal to a higher sensitivity.
We made four-point maps (offsets of 25\arcsec\ in RA and dec around the 
central pointing)
around some of the targets in order to distinguish extended structures 
and search
for cores offset from the assumed position.  In the interests of minimizing
observing time, this was done for only a subsample (62) of the cores.

We detected signal in \nh in  62\%  of our targets at the central
position, with 
the rate rising to 84\%  for the SCUBA-selected targets. 
Since \nh is a `late time' {\hk molecular ion}, not attaining significant
abundance until after $\sim10^{5}$~yrs \citep{Aikawa03}, this implies 
most of the SCUBA sources are at least this old.

Virtually every target had a \co detection (96\%).  All but one of the
spectra with signal in \nh had a corresponding detection in \cons.
Examination of the one rogue case revealed that higher than average 
noise in the \co spectrum was likely responsible for the lack of a detection.  

Table~\ref{table_detect}
shows the full break-down of the detection rates and numbers for the various 
target selection methods.  There are several factors which likely go
into the vast difference in success rates for the different selection
criteria.  Both the 2MASS and Palomar plate-selected targets have greater
uncertainties attached to a potential core's position than the SCUBA-selected
cores.  The 2MASS map has a resolution of 2.5\arcmin\ and thus is 
insensitive to the small density peak of an individual core, rather, it 
represents a smoothed average of any large- plus small-scale dense 
structures along the line of sight.  The Palomar plates
have better resolution (1\arcsec) but are only sensitive to a few 
magnitudes of visual extinction.  Our detection efficiency for this
portion of the survey is similar
to the NH$_{3}$ survey of \citet{Benson89} which used the original Palomar
plates to identify candidate cores.

\nh has a critical density of $\sim 10^{5}$~cm$^{-3}$ \citep{Tafalla02} and thus
is sensitive to only the densest gas within molecular clouds.
The SCUBA-selected and Palomar-selected targets were all chosen on the
basis of apparent (column) density enhancements which would indicate
the site of a dense core.  While this was not the case for the
2MASS-selected targets (the resolution allowed only for the identification
of peaks in the large-scale structure), those 2MASS-selected targets
which had detections all lie on or close to structure visible in the
SCUBA observations.  Hence all of the targets where we detected \nh
are likely to be dense cores rather than less dense gas unassociated with
any small-scale structure.

\subsection{Fitting the Spectra}
We reduced the \nh and \co data from IRAM using CLASS \footnote{See 
{\tt http://www.iram.fr/IRAMFR/GILDAS}.}.
First, we fit a baseline to each spectrum individually using a 4th
order polynomial.  The resultant spectra were then folded and summed
where multiple pointings existed.

To fit the seven components of the \nhns(1-0) spectra, we used CLASS's 
HyperFine Split fitting routine, with the relative frequencies and
optical depths for \nh taken from \citet{Caselli95} and 
a frequency of 93176.258~MHz for the \nhns(JF$_{1}$F=101-012) `isolated'
component \citep{Lee01}. 

In some cases (20 \nh spectra), two separate components were clearly required
for a good fit.  We interpret these spectra as belonging to two separate
entities, rather than a central dip caused by self-absorption.  
Appendix~B discusses the evidence for this interpretation {\hk of} these cores.

Table~\ref{table_fits_nh} shows the best-fit line parameters
found using CLASS for the \nh spectra -- the centroid velocity, 
velocity dispersion (measured in terms of FWHM), 
the total optical depth, the baseline level (alternatively, the standard 
deviation in regions with no line emission), and the line rms 
(alternatively the standard
deviation in the fitting residuals where there is line emission).
All fits were visually inspected;
those of poor reliability due to low S/N are noted in the final
column -- these have reasonable centroid velocities
but poor dispersions since noise limits the determination of the
extent of the line.  We include the
less secure fits only in the analysis of centroid velocities.
The integrated intensities were also measured for each spectrum
using the tdv function{\footnote {\hk The tdv function,
a part of the spectral cube package, calculates the integrated intensity 
of individual spectra between two user-specified velocities,
similar to the print area command. }}
in CLASS.  We integrated over a range of
-9 to +8~km/s around the centroid velocity fit (in order to
include all hyperfine components), and take the error in the
integrated intensity to be $B \times \delta V / \sqrt{N}$,
where B is the spectrum's rms (baseline level), $\delta V$ 
is the velocity range (17~km~s$^{-1}$) and N the number of
spectral channels summed over (541).

\begin{figure}
\plotone{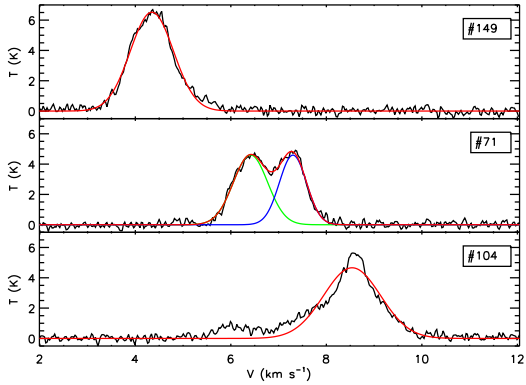}
  \caption{Three example \co spectra showing differing spectral
	profiles.  Black indicates the data while the red
	indicates the model fit.  Blue and green indicate the
	components of a two Gaussian model.  The vertical axis
	is in units of T$_{A}^{*}$.}

  \label{fig_sample_co}
\end{figure}

To fit the single-line \cons(2-1) spectra, we used CLASS's
Gaussian fitting routine, and assumed a rest frequency of 219560.354~MHz 
\citep{Muller01}.  These spectra often had complex shapes not
well approximated by a single Gaussian.  \co traces less dense gas than \nh
and thus could be expected to more often trace multiple structures along
the line of sight.  The Gaussian profile, however, is a
simple approximation for the lines and provides a rough estimate
on relevant properties.  In cases where fitting a second Gaussian made a 
marked improvement to the fit, we did so (66 spectra).  We fit a third
component in only one case (a cross position) where three distinct and 
separate features were visible.  Figure~\ref{fig_sample_co} shows
three \co spectra as an example of some of the types of profiles
observed -- the first displays an obvious single Gaussian
profile, the second a clear double-Gaussian profile and the third
a more complicated structure which we fit with a single Gaussian.

Table~\ref{table_fits_co} shows the best-fit parameters found by CLASS
for the \co spectra - the centroid velocity, velocity dispersion 
(measured in FWHM units), integrated intensity, peak intensity, baseline (the
standard deviation where there is no line emission), and line rms 
(the standard deviation of the fitting residuals where there is
line emission).  Similarly to our procedure for the \nh lines, we 
take the error in the integrated intensity
(not included in the table) to be $B \times \delta V / \sqrt{N}$
where B is the baseline, $\delta V$ is twice the FWHM and N
the number of spectral channels summed over (each spectral channel
is 0.0267~km~s$^{-1}$). 

\subsection{Other Considerations -- Pointing Accuracy}
In the analysis below, we utilize only the measurements from the
central pointing and ignore the offset cross pointings that we have on
a subsample of the targets in order to treat the entire
sample consistently.  The observations
of cross-positions are useful in allowing us to determine how accurately
we determined the \nh core centres during target selection and how much error
or bias might be introduced to our results from using only a single
pointing.  We leave a full discussion of these issues for Appendix~C,
but note the result that we find little evidence our results are
affected by any errors or bias introduced by using a single pointing.

\section{NON-THERMAL MOTIONS WITHIN DENSE \nh CORES}

The one-dimensional thermal velocity dispersion expected for \nh is given by
\begin{equation}
\sigma_{T,n} = \sqrt{\frac{k_{B}~T}{\mu_{n}~m_{H}}}
\end{equation}
where $k_{B}$ is the Boltzmann constant, T is the temperature, $\mu_{n}$
is the molecular weight of \nh in atomic units (29) and $m_{H}$ the 
mass of a hydrogen atom.  We assume a temperature of 15~K, which lies
within the range of temperatures we derived from Bonnor-Ebert modelling
of SCUBA cores in the Perseus molecular cloud \citepalias{Kirk06}.
This corresponds to a sound speed of 0.065~km/s for \nhns.
The non-thermal component of the velocity dispersion is then given
by \begin{equation}
\sigma_{NT, n} = \sqrt{\sigma_{obs,n}^{2} - \sigma_{T,n}^{2} }
\end{equation}
where $\sigma_{obs,n}$ is the observed velocity dispersion.
{\hk The level of internal turbulence, the ratio of non-thermal velocity
dispersion to the mean thermal velocity dispersion of the gas, is then :}
\begin{equation}
f_{turb} = \frac{\sigma_{NT,n}}{c_{s}}
\end{equation}
where we take a mean molecular weight of 2.33, which yields a sound
speed of 0.23~km/s in the mean gas.  

We use the same procedure to calculate the level of non-thermal
motions observed in our \co data, with $\mu_{C}$ = 30 and a 
corresponding sound speed of 0.064~km/s.

The error in $f_{turb}$ is small for either line since the linewidths
are determined to better than 5$\%$ in most cases.

Recent ammonia observations (Rosolowsky et al, in prep), which include 
all of our targets, show
a spread in temperatures between 10 and 15~K, with the mean near 12~K.
The non-thermal motions we measure would only become $\sim$10\% larger
and the turbulent fraction $\sim$10 - 35\% larger with the adoption of 
the ammonia temperatures.  We maintain our use of
15~K for consistency with \citetalias{Kirk06}.

Maps of dense cores show that they are surrounded by a less dense
envelope.  In targets where we detected an \nh signal, we expect
a surrounding envelope to also exist.  Due to chemical effects,
\co is expected to trace a larger scale than \nh even with its
smaller beamsize due to the higher frequency of the transition --
\co \citep[with critical density $\sim 10^{3}$~cm$^{-3}$, 
e.g.,][]{Ungerechts97} freezes out at densities of 10$^{5}$~cm$^{-3}$
where \nh is detectable \citep[e.g.,][]{Tafalla02}.  Therefore,
in a dense core, \co measures are weighted to the outer parts
while \nh measures are weighted to the denser inner parts.
Hence in targets where we detect both \nh and \cons, the \co can be
thought of as tracing the envelope of the dense \nh core.  
From our single pointing observations, we are unable to determine
whether this surrounding less dense gas is found distinctly around
each dense core (each core has a unique envelope) or on a larger
scale (several cores sharing a common envelope).  We therefore use
the term envelope broadly in our discussion of results.  In targets
where we only detect \cons, we do not have sufficient information
to determine if the emission originates from an envelope-like 
region.

\begin{figure}
\plotone{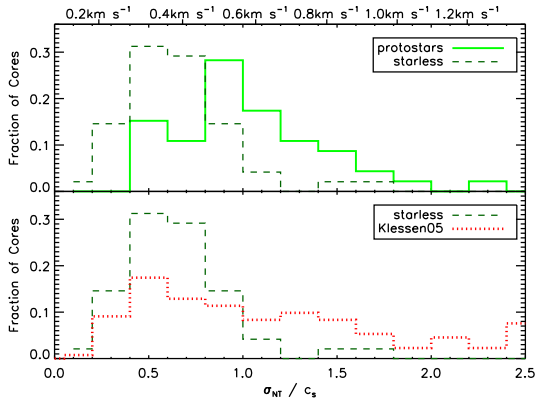}
  \caption{Relative level of non-thermal motions within the dense
	\nh cores.  The top horizontal axis shows the observed FWHM 
	linewidth in km/s while the bottom axis shows the turbulent fraction 
	f$_{turb}$ assuming a temperature of 15~K.
	The top panel shows the protostars (solid green
	line) and starless cores (dashed green line).  The bottom panel
	shows the starless cores (dashed green line) versus the
	prediction from a gravoturbulent simulation by \citet{Klessen05} for
	starless cores (dotted red line).
	Note that the final histogram bin for the \citet{Klessen05} model 
	includes all objects above this turbulent fraction (which extends
	to 4.3 in their model). 
        }

  \label{fig_internal_turb_nh}
\end{figure}

{\hk Previous observations of dense cores and their surroundings have
shown that the dense gas (observed in \nh or NH$_{3}$) traces a `coherent
core' with a close-to-constant velocity dispersion of slightly above the
thermal value \citep{Benson89,Barranco98,Goodman98,Jijina99, Caselli02}.  
Additionally, the dense core appears kinematically distinct from
the surrounding less dense gas (traced by OH or C$^{18}$O) which
displays an increasing velocity dispersion with size 
\citep{Barranco98,Goodman98}.  We therefore expect that the velocity 
dispersion we measure in our \nh pointed observations represents
the value that would be present across the entire coherence length of
the core \citep[$\sim$0.1~pc for low-mass isolated cores in][]{Goodman98}.
Figure~\ref{fig_internal_turb_nh} which plots the distribution of the level
of internal turbulence ($f_{turb}$) measured in \nhns, shows that indeed
most of the dense cores have little non-thermal motion.}

Any dense cores which have evolved to the protostellar phase might be
expected to display a greater fraction of non-thermal motions -- 
either infall our outflow motion would be expected to broaden the
line width observed.  We analyze the subset of dense \nh cores which
are not associated with protostars from the Spitzer catalog (also
plotted in Figure~\ref{fig_internal_turb_nh}) and find that, as expected, 
this subset does tend to display even less turbulent motions.
The mean and standard deviation of the turbulent fraction is
0.6$\pm$0.3 and 1.0$\pm$0.4 for the starless cores and protostellar 
cores respectively.

The turbulent fractions we find for starless cores and protostars 
are consistent with previous dense core surveys
\citep[e.g.][]{Benson89,Jijina99}.  While \citet{Jijina99} do observe
higher turbulent fractions in cores belonging to more massive and 
turbulent molecular clouds such as Orion, the turbulent fractions
we observe are consistent with the range \citet{Jijina99} find in
cores in molecular clouds with properties similar to Perseus.

The results are in contrast, however, to the
simulation of \citet{Klessen05} who find that their large-scale-driven
turbulence model only has $\sim$50\% of the cores with $f_{turb} \le 1$;
the small-scale-driven simulation has a much smaller fraction again.
The distribution of cores in the \citet{Klessen05} simulation are also
plotted in Figure~\ref{fig_internal_turb_nh}, displaying a significant
tail to the distribution of $f_{turb}$ of cores not found in our 
observations.

The turbulent fraction is also affected by association with young
stellar clusters.  \citet{Caselli95b} analyzed ammonia cores in the
Orion~B molecular cloud and found an inverse relationship between
core linewidth and distance to the nearest stellar cluster.  There
are three young star clusters near the Perseus molecular cloud --
in NGC1333, IC348, and the Perseus OB association
\citep[e.g.][]{Hatchell05}.  We note our observations are consistent
with a similar trend -- the most turbulent cores are found in
NGC1333 and IC348 (the Perseus OB association lies farther from the
cores we observed).

\begin{figure}
\plotone{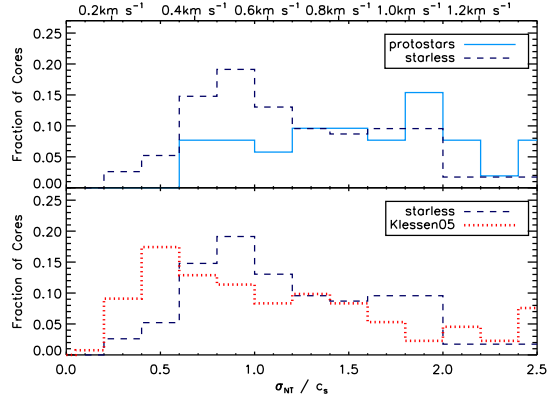}
  \caption{Relative level of non-thermal motions measured in \cons.
	The top horizontal axis shows the observed FWHM linewidth
	in km/s while the bottom axis shows the turbulent fraction 
	f$_{turb}$ assuming a temperature of 15~K.
	The top panel shows the targets associated with protostars
	(solid blue line) and those not associated with protostars (dashed
	blue line).  The bottom panel shows the targets not associated
	with protostars (dashed blue line)
	versus the prediction from a gravoturbulent simulation by 
	\citet{Klessen05} (dotted red line).
	Note that the final histogram bin for the \citet{Klessen05} model 
	includes all objects above this turbulent fraction (which extends
	to 4.3 in their model). 
	}

  \label{fig_internal_turb_co}
\end{figure}

We can perform a similar turbulent fraction analysis with the 
\co observations (adapting
the velocity dispersion equations above).  The \co observations are 
sensitive to a lower density regime and so trace larger structures
which display a higher level of turbulent motion than the 
densest parts of the core displays (see Figure~\ref{fig_internal_turb_co}).  
This is also consistent with previous observations 
\citep[e.g.,][]{Benson89}.
The results from the \citet{Klessen05} model are also plotted for
reference.  Although the model is a much closer match to these observations,
the model `observations' were designed to match to a dense-gas tracer
such as \nhns.  The core boundaries used to define the `observable area' of
the cores in the simulation, the half-maximum column density contour
\citep{Klessen05} 
better match the extent of densities traced by \nh than by \cons.
It is interesting to note that our \co targets show much less variation
in their distribution of f$_{turb}$ between those which are and are not 
associated with a protostar than the dense \nh cores do.  The mean
and standard deviation are 1.2$\pm$0.6 and 1.8$\pm$0.7 for pointings
not associated and associated with protostars respectively.
This may indicate that at the early stages of protostellar evolution, 
protostars do not affect the bulk of their envelopes in 
a significant manner!  

\section{CORE VERSUS ENVELOPE MOTIONS}

As discussed earlier, for the pointings which have both \nh and \co
detections, the \nh traces the dynamics of the dense core,
while the \co traces the dynamics of the surrounding less dense gas.

Previous studies using maps of \nh and \co have shown that cores
do not move ballistically within their envelopes (i.e. centroid
velocity differences between \nh and \co are smaller than the
\co linewidth) and most have subsonic core-to-envelope motions 
\citep[e.g.,][]{Walsh04, Walsh06}.  The \citet{Walsh04} survey examined
mostly single isolated cores, rather than those in clustered regions, while the 
\citet{Walsh06} survey spanned a clustered region of dense cores 
(NGC1333, which is included in our larger-area sample
although at lower resolution).  Unfortunately, the \citet{Walsh06} 
results are less certain since their \co beamsize was significantly 
larger than their \nh beamsize.
Thus our survey is ideal in providing a large statistical measure
of core-envelope motions within a clustered environment.  

\begin{figure}
\plotone{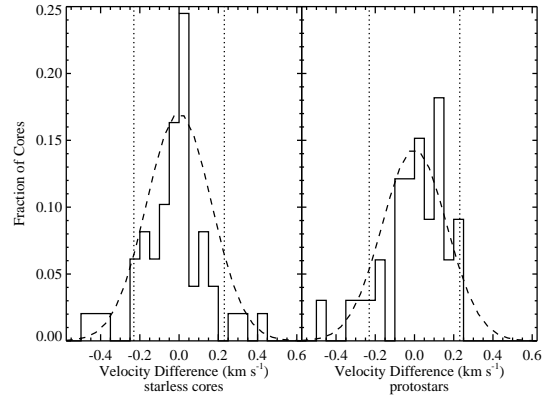}
  \caption{Difference in centroid veloicites of \nh to \co 
	for the starless cores (left) and protostars (right).  The
	dotted lines indicate the sound speed of the ambient medium.
	The dashed lines indicate Gaussian fits to the distributions 
	-- the starless cores have $\sigma =$ 0.17~km/s while the
	protostars have $\sigma =$ 0.16~km/s.}
  \label{fig_core_rel_envelope_1}
\end{figure}

We separately analyze the relative core to envelope motions of the starless 
and protostellar cores.  In starless cores, the relative motions
should be induced by the molecular cloud and the core formation
process.  In protostellar cores, the relative motions between
core and envelope could be complicated by outflows or processes which
decouple the core from its envelope.  Figure~\ref{fig_core_rel_envelope_1}
shows our results for the starless cores and protostars separately --
both have a high fraction of relative velocities which are less
than the sound speed.

\citet{Ayliffe06} has argued that the previous analyses of
\citet{Walsh04,Walsh06} biases results towards small velocity
differences due to the method of analysis.  In instances where
multiple CO velocities were found along the line of sight, 
the velocity component closest to the \nh core velocity was assumed
to be the one associated with the core.  Here, we demonstrate that
taking the closest CO velocity component is reasonable and does 
not introduce significant
bias.  Most of the cores in our observations had only a single
velocity component fit.  For these, we compared the difference in
centroid velocity between the \nh and \cons, as shown in 
Figure~\ref{fig_core_rel_envelope_1}.  This figure 
demonstrates that the vast majority of cores (nearly 90\%) have differences
less than the sound speed of the ambient medium (dotted lines), and
the remaining cores have differences which are not much larger.  
Note that some of the \nh cores
in this plot were fit to two velocity components; we considered
these two velocity components as separate entities (i.e., plotted
as two distinct cores).  The velocity differences found for the
two \nh velocity cores versus the surrounding \co tend to be
larger than for the other cores (since both \nh velocities are 
compared to the same \co velocity).  The CO linewidth for the two
velocity \nh cores also tends to be broader than average.
All of the cores have an error in the centroid velocity of
around several hundredths of a km~s$^{-1}$, indicating that the majority
of the differences in velocities observed are real.

\begin{figure}
\plotone{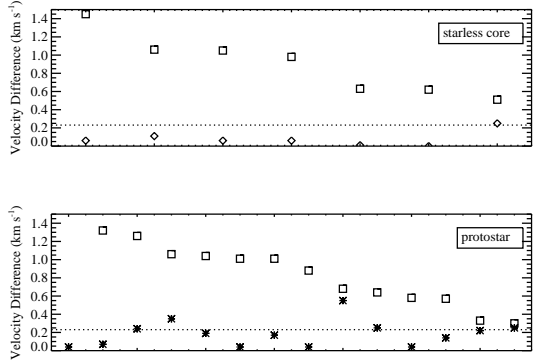}
  \caption{Difference in centroid velocities of \nh and \co for
	starless cores (top) and protostars (bottom).  The closest \co
	velocity components are denoted by diamonds or asterisks while 
	the farther
	\co velocity components are denoted by squares.  In each instance,
	the cores are ordered from largest to smallest difference in 
	velocity using the farther \co velocity component.  The dotted
	line indicates the sound speed in the ambient medium.}
  \label{fig_core_rel_envelope_2}
\end{figure}

Figure~\ref{fig_core_rel_envelope_2} shows the absolute difference
in \nh to \co centroid velocity in dense cores where
two \co velocity components were fit, again split into starless cores
and protostars.  The velocity differences here are ordered in terms of 
the largest-difference
component (squares), while the smallest-difference component 
is denoted by diamonds or asterisks.  Clearly the two CO velocity 
components are not correlated, as is expected since the second 
CO component merely lies along the same line of sight.  
The cores which happen to have a second \co velocity
component along the same line of sight should possess a similar distribution
of core-to-envelope relative velocities as 
along lines of sight where only a single \co velocity component
was observed -- i.e., velocity differences smaller than the sound speed in the 
vast majority of cases.  As can be seen from 
Figure~\ref{fig_core_rel_envelope_2}, in almost every case this implies the
closer CO velocity component is the only sensible one to associate
with the dense core; in the rare instance of ambiguity, the resultant
number of cores with each velocity difference will be little affected.

Assuming in all cases that the closest velocity \co component is the one
associated with the \nh profile peak, we find that in the majority of dense
cores, the core-to-envelope velocity tends to be smaller than thermal, 
(88\% and 83\% for the starless and protostellar cores respectively). 
In the \citet{Walsh04}
survey of mostly isolated cores, they find only one out of 35 cores
(or 3\%) with a core-to-envelope velocity exceeding the sound speed.
The \citet{Walsh06} survey of cores in NGC1333 found an rms
core-to-envelope velocity of 0.53~km/s; nearly half of their cores
have differences greater than the sound speed of the medium.  Our
survey includes the NGC1333 cores but finds much lower differences
(we have rms velocity differences of 0.16~km/s for the starless
cores and 0.18~km/s for the protostars when both the one- and two-
velocity component CO spectra are included).  This may indicate that 
the \citet{Walsh06} survey results were biased by the much larger
beamsize for \co (50\arcsec) than \nh (10\arcsec) which could have
sampled a large fraction of material not associated with the individual
dense core's envelope.

As discussed in \citet{Walsh04}, small relative motions between cores 
and envelopes could be interpreted
as an indication of quiescence on small scales, and as such would
appear to argue against a competitive accretion scenario for star
formation, where dense cores gain most of their mass by sweeping up
material as they move through the cloud.  Recent analysis of
simulations by \citet{Ayliffe06}, however, {\hk show} that the competitive 
accretion scenario is not necessarily incompatible with the observations.
\citet{Ayliffe06} analyzed
the competitive accretion simulations of \citet{Bate03} and demonstrated 
that the simulated observations also show core to envelope motions are
not ballistic.  There were, however, differences between the simulated
observations and the \citet{Walsh04} results at later times in the
simulation (such as the \nh linewidth becoming larger than the \co
linewidth), which were attributed to the clustered environment of the
simulation, versus the isolated cores observed.  Since our observations
probe cores forming in a clustered environment (and include protostars), 
we can make a stronger comparison with the \citet{Ayliffe06} results at
later times.  \citet{Ayliffe06} do find the majority of their
sources have velocity differences less than the sound speed at all
time steps, however, they have a more significant tail out to large
velocity differences (around 0.5~km/s or higher).  They find the 
dispersion in the velocity difference ranges from 0.25 to 0.27~km/s,
or 0.18 to 0.24~km/s after smoothing to the \citet{Walsh04} resolution.
It should also be 
noted that the \nh linewidth becomes equal to or larger than the
\co linewidth at 1.1 times the free-fall time and beyond, contradicting 
our observations (\S4) and that of many previous studies.
Thus while the simulation of competitive accretion analyzed by
\citet{Ayliffe06} has promise in reproducing core-to-envelope
dynamics it does not simultaneously reproduce all observations.

Our results show no indication that the core-to-envelope motions 
significantly change between the starless and protostellar stages of 
evolution.

On an even smaller scale, \citet{Jorgensen07} examined the location of
YSOs within SCUBA cores and showed that they lie within 15\arcsec of 
the SCUBA core centre.  These small separations imply that the YSOs
have motions smaller than the thermal velocity relative to the SCUBA
core they were born in.  The picture that emerges 
from the combination of these results is that the central source, core,
and envelope are quiescent.

\section{CORE-TO-CORE MOTIONS}
We next examine how the \nh cores move with respect to each other to gain an
understanding of the dynamics on larger scales within the cloud.  

We can use our extinction map of the Perseus molecular cloud to
define the larger regions in which the dense \nh cores inhabit.
In \citetalias{Kirk06}, we identify large-scale structure in the extinction map 
which we will term `extinction regions' here to prevent confusion (the 
regions are referred to as `extinction super cores' in \citetalias{Kirk06}).
Starless cores within each extinction region should be coupled to the 
surrounding gas
in the region, and hence the motion of the starless cores should reflect 
the motion of the ambient material.  Protostellar cores may have become
decoupled from their parental material, and hence are a less reliable
tracer of the dynamics occuring in the region.

We analyze the motions within each extinction region and determine
whether the regions appear to have sufficient
velocity dispersion to provide support against gravity.  We adopt the
commonly used formulation of \begin{equation}
\sigma_{grav} = \sqrt{GM_{ext}/5R_{ext}}
\end{equation}
as the velocity dispersion required in 1-D to prevent collapse 
\citep[see for example][]{Bertoldi92}.  We estimate the total mass
and size of each region from the extinction data \citepalias{Kirk06};
{\hk these} data {\hk are} provided in Table~\ref{extinct_regions}.  
The above formula technically only applies to a uniform density sphere,
but different density structures and object shapes change 
the required velocity dispersion by factors of order
unity \citep{Walsh06,Bertoldi92}.  Regions which display 
$\sigma_{obs} = \sigma_{grav}$ are often said to be in approximate
virial equilibrium, although to have true virial equilibrium, the
`surface terms' of the virial equation must be included 
\citep[e.g.,][]{Dib06}. 

\begin{figure}
\plotone{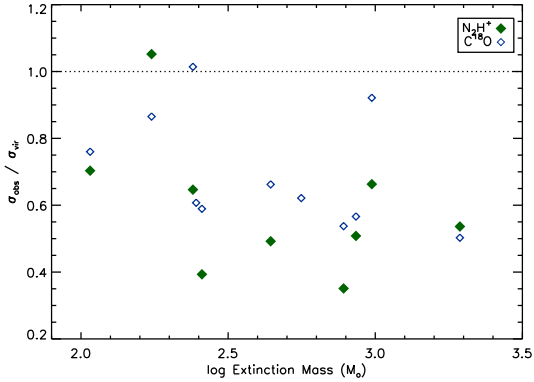}
  \caption{Ratio of the measured velocity dispersion to that required to
	counteract gravity ($\sqrt{GM_{ext}/5R_{ext}}$)
	versus the mass in the extinction region.
	The dotted line shows the expected relationship for virial equilibrium.
	The green filled diamonds denote the dispersion in centroid velocity for
	\nh starless cores.  The blue open diamonds denote the dispersion of
	the summed \co spectra for all starless cores in the extinction 
	region.  Two extinction regions have no \nh dispersion
	measured since less than two \nh cores were detected. }

  \label{fig_core_core}
\end{figure}

Motions providing support for the extinction region could
originate on either the small or large scale.  The former would
be measurable through internal core motions, while the latter 
through core-to-core motions.  In the case of the
densest material probed by \nhns, the internal core motions are of order
the thermal motions of a gas at $\sim$15~K.  If the extinction regions
were in virial equilibrium, they would require effective temperatures
of up to several hundred Kelvin to prevent gravitational collapse
\citepalias{Kirk06}; therefore, internal thermal motions cannot provide 
the bulk of the support required and hence most of the support must
originate in large scale motions.
We measure the dispersion in centroid velocities of the \nh starless
cores within each extinction region to determine the amount of support
that can be provided by large scale motions traced by the cores.  
We add the contribution of thermal motions (which has little effect
except for the smallest core-to-core velocity dispersions).  This total
support is plotted 
in Figure~\ref{fig_core_core} in terms of the ratio of observed velocity
dispersion to that which is required for virial equilibrium.
The horizontal axis plots the mass within each extinction region.

In the case of the material probed by \cons, small-scale motions could 
provide a larger contribution to overall support, since the internal 
velocity dispersions are often several times larger than the sound speed 
and are closer in magnitude to the point-to-point velocity differences.
In order to account for both of these contributions and also to
decrease potential errors from the difficulty in fitting each \co
spectrum,
we sum all of the spectra within an extinction region and fit
the sum with a single Gaussian thus measuring the total velocity
dispersion within each region.  We correct the thermal component
of the velocity dispersion to be that of the mean gas, rather than
\cons.  These results are shown in the plot as blue open diamonds.  

Table~\ref{extinct_regions} also summarizes the relevant information for
each extinction region -- the number of non-protostellar cores detected 
in \nh and \co in each region and the velocity dispersions
measured with both molecules.

Variations by a factor on the order of one could be expected between the
estimated and true velocity dispersion required for gravitational
support -- the extinction regions do not have a spherical geometry,
several extinction regions have a small number of cores to calculate the
velocity dispersion from, the cores do not span the entire extent 
of the extinction regions,
and the conversion between extinction and mass has some uncertainty.
It should also be noted that there is a velocity gradient across the
Perseus molecular cloud -- we leave a detailed analysis of the
core motions relative to the overall cloud gradient for a
future paper, and do not attempt to correct for it when calculating
the velocity dispersions used here.
With these considerations in mind, we find that the extinction
regions tend to display velocity dispersions lower than required for
`virial equilibrium', with the starless cores possessing lower 
dispersions in \nh than in \cons, but the measurements do not rule
out `virial equilibrium'.  

\section{ENVIRONMENTAL EFFECT ON DENSE CORES}
{\hk Most (84\%) of the SCUBA cores have associated \nhns.  For these 
cores, we can examine whether the SCUBA core properties have an 
effect on their internal dynamics.  Since \nh requires 
$\sim 10^{5}$ years to form \citep{Aikawa03}, the SCUBA cores must be
at least this old, and hence are at an advanced stage of
evolution, consistent with the results of \citet{Jorgensen07}. }

\subsection{Concentration}
\begin{figure}
\plotone{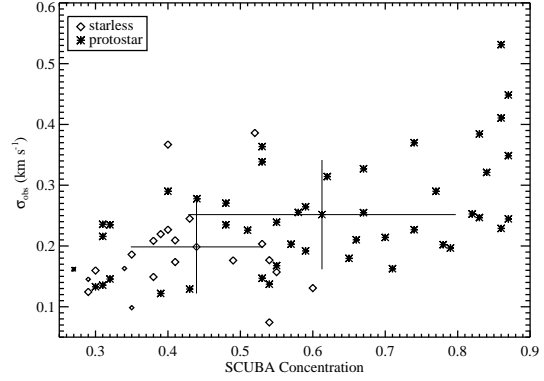}
  \caption{Variation of observed velocity dispersion in \nh cores
	versus concentration for starless cores (diamonds) and protostellar 
	cores (asterisks).  The large bold crosses indicate the
	mean and standard deviation for each of the two samples.
	The smaller symbols indicate SCUBA sources for which the properties
	derived from clumpfind are less secure (see discussion in 
	Appendix A).  These sources were not used in any of the 
	calcuations.}
  \label{fig_internal_turb_concs}
\end{figure}

The `peakiness' or central concentration of a core gives an 
indication of the importance of self-gravity of the core.
The concentration is defined in terms of observable measures
in equation (A1).
Concentration can be thought of as an approximate proxy for evolutionary
state with high concentration objects being more evolved 
\citep{Walawend05,Walawend06,Johnstone06,Jorgensen07} -- in the framework of 
a Bonnor Ebert sphere model, any object with concentration above
0.72 is unstable to gravitational collapse, furthermore, heating from
a central source also leads to an increase in concentration. 
Figure~\ref{fig_internal_turb_concs}
shows the variation in observed core velocity dispersion with SCUBA 
concentration
for both protostellar and starless cores.  The mean concentration and 
velocity dispersion
are lower in the starless cores than the protostars,
with protostars of higher velocity dispersion also possessing high
concentrations.
The mean and standard deviation of the concentration is C = 0.4$\pm$0.1
and C = 0.6$\pm$0.2 for the starless and protostellar cores respectively.
The velocity dispersion observed for cores associated with a SCUBA source
is 0.20$\pm$0.08~km/s for the starless cores and 0.25$\pm$0.09~km/s for the 
protostars.
The scatter in the velocity dispersion of the starless cores is mostly
due to two starless cores with unusually high internal 
turbulence levels ($>$ 1.5); excluding these two cores, the velocity
dispersion becomes 0.18$\pm$0.04~km/s.  These
two starless cores are in NGC1333 where the region is highly clustered,
making it more difficult to determine accurate core properties as
well as determine if there is an associated protostar.
The velocity dispersion for those cores which are not associated
with a SCUBA source (primarily targets selected from the Palomar
plates) tend to be even lower -- the mean and standard
deviation is 0.14$\pm$0.04~km/s.

\subsection{Total Flux}
\begin{figure}
\plotone{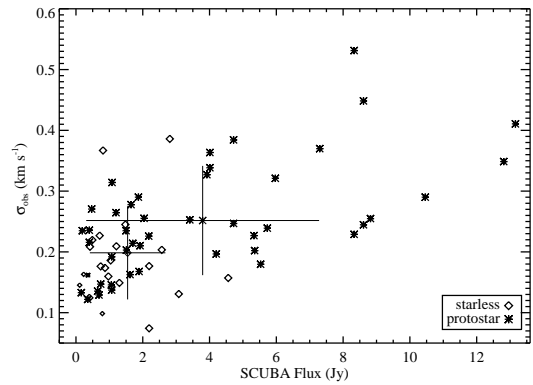}
  \caption{Variation of core velocity dispersion with the
	total flux within a SCUBA core for both protostars and
	starless cores.  The same plotting convention is used
	as in Figure~\ref{fig_internal_turb_concs}.
	}
  \label{fig_internal_turb_flux}
\end{figure}

We next examine the relationship between internal turbulence
level with the total flux of the SCUBA core.
Figure~\ref{fig_internal_turb_flux} shows SCUBA core total flux versus
the velocity dispersion for both the starless and protostellar 
cores.  No trend is apparent for the starless cores (diamonds), while
there appears to be a weak trend of higher flux corresponding to higher 
velocity dispersion in the protostars (asterisks).  If we split the protostars
into those with fluxes greater than 5~Jy and less than 5~Jy,
we find the mean and standard deviation are 0.31$\pm$0.10~km/s and 
0.23$\pm$0.07~km/s respectively.  The {\hk symbols intersected by
crosses} show the mean and standard deviation for the
protostars and starless cores, 
indicating that the protostars tend to have higher flux than the 
starless cores.  This could be the result of slightly higher central 
temperatures in the protostellar cores.

If we assume a constant temperature of 15~K, a dust opacity of 
0.02~g$^{-1}$cm$^{2}$ at 850~$\mu m$ and a distance to the
Perseus molecular cloud of 250~pc \citep[e.g.,][]{Cernis93}, we can convert
the observed SCUBA flux into mass as 1~Jy = 0.48~M$_{\odot}$ 
\citepalias[see][]{Kirk06}.  Note that due to the non-negligible 
uncertainties in all of 
the above quantities, the mass is only accurate to a factor of 
roughly 6.  Even with the large uncertainty, 
we can use the mass and radius measured for each SCUBA core to
estimate the internal velocity dispersion required to provide
support against gravity, which we again take to be 
$\sigma_{grav} = \sqrt{GM_{C}/5R_{C}}$ (c.f. eq. [4]).

\begin{figure}
\plotone{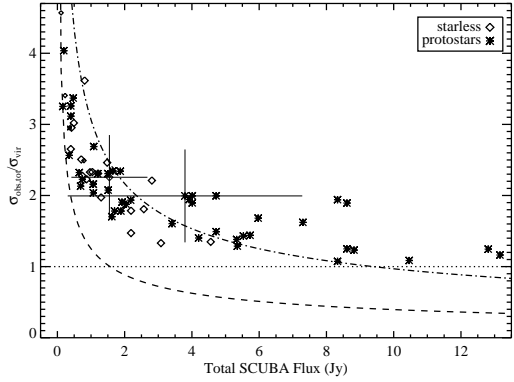}
  \caption{Ratio of the mean gas velocity dispersion to the
	virial velocity dispersion (calculated from SCUBA core flux and
	radius) squared, versus the observed SCUBA flux.  The same plotting
	conventions are used as in Figure~\ref{fig_internal_turb_concs}.
	The dotted line shows the expected relationship for virial equilibrium.
	The dashed and dot-dashed lines show the relationship for thermal
	15~K cores assuming core radii of 10\arcsec\ and 60\arcsec\ 
	respectively.} 
  \label{fig_internal_turb_mass}
\end{figure}

The velocity dispersion of the mean gas can be calculated by correcting
for a thermal component with a molecular weight of 2.33 rather than
29 for \nhns.  We can then compare the total gas velocity dispersion to
that predicted for virial equilibrium (from the SCUBA observations).
Figure~\ref{fig_internal_turb_mass} shows the square of the ratio of the 
observed velocity dispersion to the virial velocity versus SCUBA flux 
for the starless cores and protostars. `Virial equilibrium' would
occur for a ratio of 1 (dotted line).  All of the \nh cores have
higher velocity dispersions than predicted by the virial equation,
with those at small SCUBA fluxes displaying the largest difference.
The cores farthest
from `virial equilibrium' would require the SCUBA mass to be underestimated by
a factor of ten or more if they were truly in virial equilibrium, far
larger than our uncertainties allow.  Many of the cores would be far from
virial equilibrium even if their observed velocity dispersion were purely
thermal -- the dashed and dot-dashed
lines show the relationships for a 15~K thermal core velocity dispersion 
and assuming core radii of 10\arcsec\ and 60\arcsec\ respectively 
(bounding the observed range of SCUBA core radii).

The above analysis ignores the contribution of external pressure
in the virial equilbrium calculation which for sub-Jeans mass objects
keeps the internal motions thermal even though gravity alone does not
require internal motions of this magnitude.  {\hk In \citetalias{Kirk06}, we} 
find that
the SCUBA cores in Perseus are well fit by Bonnor-Ebert spheres with
external pressures in the range of 5.5~$\le$~log$_{10}$P$_{ext}$/k$_{B}$
$\le$~6.0.
For a `typical' core of roughly one solar mass and 50\arcsec\ in extent,
when the external pressure is included in calculating virial equilibrium,
the square of the virial velocity rises by a factor of approximately 1.3 
to 2 of what it was without considering the external pressure, which 
would make most of
the higher flux cores in approximate virial equilibrium.  
Similarly, a critical BE sphere, which has
R$_{crit}$=0.41$\frac{GM}{c_{s}^{2}}$ \citep{Hartmann98}, 
requires $\sqrt{5\times0.41}$, or roughly 1.4, times the velocity
dispersion one would naively assume. 
Note that observing a line width and converting to mass using the virial
equation without accounting for surface pressure will overestimate the
enclosed mass of an equilibrium core.  See also \citet{Dib06}.

Our results thus show that while we observe velocity dispersions
that are several times larger than what is predicted by the
traditional `virial equilibrium' measures, when the external
pressure on the dense cores from the ambient cloud is accounted
for, the agreement is reasonable for the higher flux cores.
This is in contrast with some previous observations which
tend to find velocity dispersions which are consistent
with `virial equilibrium' without accounting for any
external pressure -- for example, the low mass dense core
survey of \citet{Caselli02}.  Some previous studies, however, 
have shown that external pressure is required for virial equilibrium, 
e.g., in the Horsehead nebula \citep{Wardthomp06}.

The turbulent simulations of \citet{Klessen05} predict a 
relationship between virial and observed mass -- 
their large-scale-driven simulation (which more closely matches our
other observations) shows that the starless cores have
virial masses which are greater by up to a factor of thirty 
than the actual mass.  Unlike our observations, however,
the simulation shows that protostars have virial masses which
are several times less than the actual mass.  
\citet{Klessen05} points out that the virial mass estimates
for protostars are underestimated due to the lack of
velocity resolution of the gas in the central sink cell, but that
this will have a small effect on the measured velocity dispersion
since a small fraction of the core mass is contained within the
sink cell.

\subsection{Variation of Line Intensity}

\nh and \co observations can also serve as a probe of the chemistry
of the dense cores.  \nh is only able to form
in significant amounts after \co freezeout has occurred, as the two
molecules form via competing reactions.  While \nh is observed to be a good
dense gas tracer for densities of $10^{5} - 10^{6}$~cm$^{-3}$
\citep{Tafalla02}, it may freeze out onto dust grains at densities 
above this \citep{Crapsi05}.  \co on the other hand, should be depleted
at high densities.  
At later stages of evolution once a central protostar has formed,
the situation is expected to reverse, with the central region heating, 
causing the liberation of CO and destruction of \nhns.
Our SCUBA observations allow us to estimate the (column) density of 
the dense cores independently of our IRAM observations which may 
be affected by chemistry.  Using the
same flux to mass conversion factors discussed above, we can
convert the SCUBA flux into a column density --
1~Jy~beam$^{-1}$ corresponds to 0.24~g~cm$^{-2}$ or $\sim10^{22}$~cm$^{-2}$.  
Making the further assumption
that the cores are roughly spherical and have a diameter of $\sim$~50\arcsec\
in the plane of the sky, this corresponds to a density of 
$\sim 10^{6}$~cm$^{-3}$.

\begin{figure}
\plotone{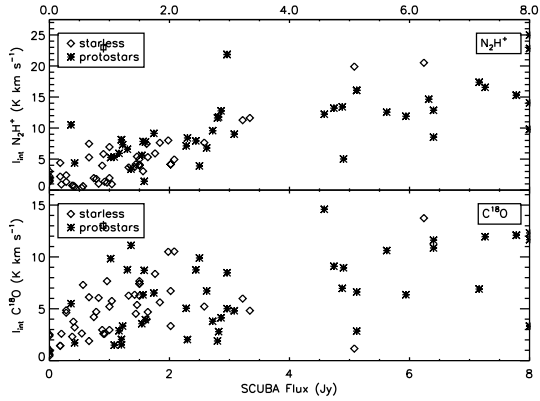}
  \caption{Variation in integrated intensity for \co and \nh
	with total SCUBA flux over the region observed by IRAM.  
	Note that cores associated with total
	SCUBA fluxes of over 8~Jy were included in the plot
	as having SCUBA fluxes of 8~Jy~beam$^{-1}$.  The same plotting
	conventions are used as in Figure~\ref{fig_internal_turb_concs}.
	The squares with vertical lines indicate the mean size of the
	error in the integrated intensity.  The error bar for the
	\co integrated intensity has been enlarged by a factor of two.
	}
  \label{fig_intens_vs_submm}
\end{figure}

Figure~\ref{fig_intens_vs_submm} shows the total SCUBA flux within the
IRAM beam for each observation versus the integrated intensity measured 
in both \nh and \cons.  The \nh integrated intensity shows some correlation
with the total SCUBA flux for both the starless and protostellar cores. 
The \co integrated intensity possibly shows a very weak correlation with
SCUBA flux below $\sim$1Jy and no correlation above this.  This is consistent
with denser cores being dominated by central freeze out 
(i.e., even with increasing column density, the 
\co integrated intensity remains constant). 

\begin{figure}
\plotone{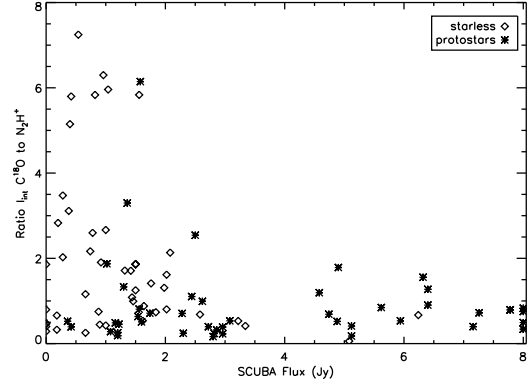}
  \caption{Ratio in \co to \nh integrated intensity versus total SCUBA
	flux over the IRAM beam.  Note that all cores associated with 
	SCUBA fluxes of over 8~Jy have been included in the plot as 
	having values of 8~Jy.  The diamonds denote starless cores while 
	the asterisks denote the protostars.}
  \label{fig_intens_ratio_vs_submm}
\end{figure}

We also examine the
ratio between the \co and \nh integrated intensities -- a low ratio, for 
example, would be indicative of freezeout.
Figure~\ref{fig_intens_ratio_vs_submm} shows the total SCUBA flux
observed in each IRAM beam versus the ratio of integrated intensity measured
in \co and \nhns.  This figure shows that high \co to \nh ratios mostly
occur for starless cores, and only at smaller SCUBA fluxes, i.e., where 
the density is lowest and hence there is little to no freeze out.  
All high flux SCUBA cores have low \co to \nh ratios.  Due to
the large relative error in the \nh integrated intensity for these
cores, the error in the ratio is often greater than 100\%.
Note that following
the results of \S5, in the few cases where two \co components were
associated with an \nh dense core, the integrated intensity of the
component with the closest velocity is plotted.

We can also calculate the \nh column density from the integrated intensity,
assuming an excitation temperature of 15~K and correcting for the optical
depth.  We use eq. (10) of \citet{Shirley05}, and find a column density of
\begin{equation}
N_{N_{2}H^{+}} = 1.47\times10^{7} \frac{\langle \tau \rangle}
{1-e^{-\langle \tau \rangle}} 
\int T_{A}^{*} dV \times \frac{F_{eff}}{B_{eff}}\  cm^{-2}
\end{equation}
where $\langle \tau \rangle$ is the mean optical depth of the hyperfine
transitions (CLASS's hfs fitting routine fits 
$\tau_{tot} = 7 \langle \tau \rangle$), $\int T_{A}^{*} dV$ is the
integrated intensity in K~m~s$^{-1}$, and $F_{eff} = 0.95$ and 
$B_{eff} = 0.77$ are
beam efficiency parameters available from the IRAM 30m website.
Our minimum observable column density is $\sim 10^{11}$~cm$^{-2}$.

\begin{figure}
\plotone{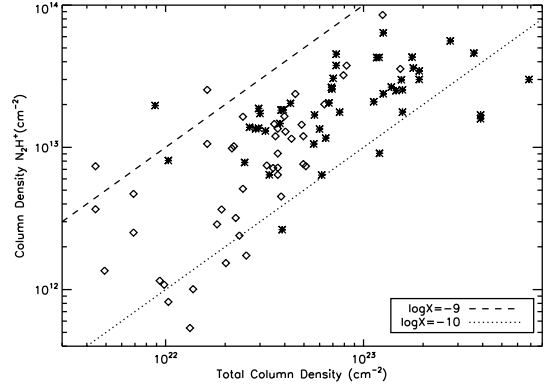}
  \caption{Column density of \nh versus the total column density 
	(calculated using the SCUBA flux within the IRAM beam).
	The diagonal lines indicate fractional abundances of \nhns. 
	Starless cores are diamonds while protostars are asterisks.}
\label{fig_coldens}
\end{figure}

Figure~\ref{fig_coldens} shows the \nh column density derived versus
the total column density derived from the total SCUBA flux measured in the
IRAM beam.  The relative error in the total column densities is
$\sim$30\% (the calibration error of SCUBA data), with an absolute error
close to a factor of six due to uncertainties in constants used to convert
flux to mass.  The errors in the \nh column density vary substantially,
mostly due to errors in the optical depth determined; the median error is
30\%.  The noise in the SCUBA map is $\sim$10~mJy~beam$^{-1}$, which 
translates to
a minimum observable total column density of $\sim 10^{21}$~cm$^{-2}$.

Overplotted on Figure~\ref{fig_coldens} are lines of constant \nh abundance.
The cores lie between approximately N$_{N_{2}H^{+}}$ / N$_{H_{2}} = 10^{-9}$ and
10$^{-10}$, 
consistent with what has been more accurately derived from detailed mapping
and analysis of single cores \citep[e.g.,][]{Shirley05,Tafalla04}.

\section{CONCLUSIONS}
We present results from a survey of \nhns(1-0) and \cons(2-1) of 157 dense 
core candidates in the Perseus molecular cloud.  We detect \nh in
62\% of our targets, and 84\% of our SCUBA-selected targets.  
\nh is a `late-time' {\hk molecular ion} which does not become abundant until
$\sim10^{5}$~years \citep{Aikawa03}.  Since we detect \nh in the vast majority
of SCUBA cores, this argues that objects which attain sufficient
density to be detectable with SCUBA are not short-lived, transient
objects.  This is in agreement with the findings of \citet{Jorgensen07}
who argue that starless SCUBA cores have roughly equal lifetimes to
that of deeply embedded protostars, which is on the order of $10^{5}$ years
\citep{Wardthomp07}. 

We differentiated between starless cores and protostars
on the basis of Spitzer data \citep{Jorgensen07}.  In \nhns, the starless
cores have linewidths which are dominated by thermal broadening, while
the protostars have slightly larger linewidths, consistent
with many previous surveys including \citet{Benson89} and \citet{Jijina99}.
We find fewer \nh cores dominated by non-thermal motions than 
predicted by the turbulent simulations of \citet{Klessen05}.
For the starless cores, the mean ratio of non-thermal to thermal motions
($f_{turb} \sim 0.6$) implies the ratio of non-thermal to thermal
pressure, $f_{turb}^{2}$, is less than 40\%.  Naively, this runs
counter to turbulent models where cores as well as transient density
peaks form at the convergence of supersonic flows.  Simulations must
therefore demonstrate that turbulent pressure does not dominate
in the high density regime probed by \nhns.

The \co observations, sensitive to lower density material, reveal
much more non-thermal motions as previous surveys have also shown.  There is 
less difference in non-thermal motions between
those targets associated with protostars and those which {\hk are not} associated
with protostars
which may imply that protostars have little effect on the dynamics of bulk
of their envelopes at the earlier stages of evolution.

We find the relative motions of the dense \nh cores and their envelopes 
(measured in \cons) tends to be less than thermal in the majority of cases,
confirming and strengthening the results of \citet{Walsh04,Walsh06}
for clustered star forming environments.  

Within large scale structure, defined through 2MASS extinction observations, 
the core-to-core motions of starless
cores are not sufficient to provide support against gravity, however,
the sparse sampling of each extinction region leads to large errors
associated with the core-to-core velocity dispersions we measure.
The total velocity dispersion tends to be smaller when measured in
\nh than \co due to the smaller linewidths seen in \nhns.

The \nh cores which have an associated submillimetre source detected with 
SCUBA have internal motions several times larger than is required 
to provide support against gravity.
Inclusion of external pressure shows the cores to be in approximate
virial equilibrium.  The protostars tended to have higher SCUBA concentrations,
and total fluxes.
High ratios of \co to \nh integrated intensity, possibly indicating
chemically young gas, were found for some cores which had a low flux
measured with SCUBA.  At higher SCUBA fluxes,
only low ratios of \co to \nh integrated intensity were observed.
Column densities derived for \nh were consistent with abundance ratios
between 10$^{-9}$ and 10$^{-10}$, in agreement with what has been 
previously derived for cores with more accurate observations.

Our survey utilized single pointings on most of our dense core
candidates, rather than using the traditional route of mapping.  We
show that the lack of a full map around each source has a minimal
effect on our dynamical analysis.  High resolution maps are, however, quite
helpful in disentagling the motions in complex regions such as
NGC1333 where multiple objects along the line of sight could otherwise
lead to confusion in interpretation of results.  Our method is
an efficient and effective way to survey the dynamics of dense cores
over the full extent of a molecular cloud.  In the future, this
method can be applied to other molecular clouds in order to
determine whether the dynamical properties of the cores observed
in the Perseus molecular cloud are universal or are dependent
on the cloud environment.

\acknowledgements
We thank the IRAM 30~m staff for their hospitality and support during our 
observations.

The Second Palomar Observatory Sky Survey (POSS-II) was made by
the California Institute of Technology with funds from the National
Science Foundation, the National Geographic Society, the Sloan
Foundation, the Samuel Oschin Foundation, and the Eastman Kodak
Corporation.

HK\footnote{Guest User, Canadian Astronomy Data Centre, which is 
operated by the Herzberg Institute of Astrophysics, National Research 
Council of Canada} 
is supported by a Natural Sciences and Engineering Research Council
of Canada CGS Award and a
National Research Council of Canada GSSSP Award.  DJ is supported 
by a Natural Sciences and Engineering Research Council of Canada 
grant.

HK would like to acknowledge valuable discussions with people at the
CfA -- other members of the COMPLETE collaboration, in particular
Alyssa Goodman, Jens Kauffman, and Erik Rosolowsky, as well as 
Phil Myers and Charles Lada.  Additionally, HK thanks Matthew Bate 
for a sending a preprint of his work.

\appendix

\section{SCUBA OBSERVATIONS}
\begin{figure}
\plotone{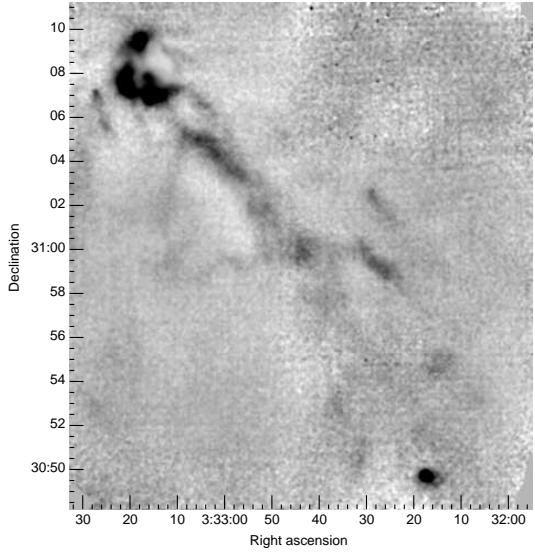}
  \caption{The 3\arcsec\ SCUBA map of the B1 star forming region.
	The image is scaled such that white corresponds to 
	$\sim$0~Jy~beam$^{-1}$and black to $\sim$0.25~Jy~beam$^{-1}$.}
\label{fig_submm_ngc1333}
\end{figure}

In this paper, we utilize newly created 850~$\mu$m SCUBA maps with a finer
sampling size of 3\arcsec\ to better determine SCUBA source properties.
We used our full 6\arcsec\ sampled map of \citetalias{Kirk06} to define
regions in which to create 3\arcsec\ sampled maps.  As in \citetalias{Kirk06},
we combined all of the scan- and jiggle- map data in the JCMT 
archive\footnote{Based on observations obtained with the James Clerk Maxwell 
Telescope, which is operated by the Joint Astronomy Centre in Hilo, Hawaii 
on behalf of the parent organizations PPARC in the United Kingdom, the 
National Research Council of Canada and The Netherlands Organization for 
Scientific Research.} 
and used the same reduction procedure as
in \citetalias{Kirk06}, making modifications only for the smaller sampling size.
We first use the normal SCUBA software \citep{Holland99} to flat-field and
atmospheric-extinction correct the raw data.  We then used the
matrix inversion technique of \citet{Johnstone00} to produce the images.
The matrix inversion techniqe has been shown to produce better
images than the standard procedure used at the JCMT, particularly
when combining data of different qualities \citep{Johnstone00}, as is the
case for the archival data used here.  In order to correct for atmospheric
fluctuations and other effects, SCUBA data is in the form of a series of 
difference measures (chops).  Any image-reconstruction technique is 
thus insensitive to real structures which have sizes several times
larger than the chop throw \citep{Johnstone00}.
We remove this structure by subtracting a large-scale
smoothed version of the map from the original (we smooth with a Gaussian
of $\sigma = $90\arcsec).  In order to prevent the introduction of 
negative `bowls' around bright sources (and similarly diffuse
positive regions around deep compact `holes'), we first create a
map where all values outside of $\pm 0.1$~Jy~beam$^{-1}$ per pixel 
were replaced with those
values before smoothing to create the large-scale smoothed map 
(0.1~Jy~beam$^{-1}$ corresponds to roughly five times the rms value).  We 
also smoothed pixel-to-pixel noise using a Gaussian with $\sigma = 3$\arcsec. 
Figure~\ref{fig_submm_ngc1333} shows our map of {\hk B1} as an example
of the 3\arcsec\ SCUBA maps.

To identify SCUBA cores in the maps, we utilized the object-identifying
algorithm of \citet{Williams94}, `Clumpfind 2D'.  In this algorithm, objects
are identified as peaks at 2~`$\sigma_{C}$' intervals and extended until 
they either encounter another object or the lowest allowed `$\sigma_{C}$'
level.  Normally, $\sigma_{C}$ is the noise level in the map, 
however, in order to have a consistent core-identification threshold
in all of the 3\arcsec\ mapped regions, we used $\sigma_{C} =$ 
0.03~Jy~beam$^{-1}$
per pixel, which corresponds to approximately the same level in which we 
identified SCUBA cores in our previous work \citetalias{Kirk06}.  Accurate 
noise levels for each 3\arcsec\ sampled map were difficult
to determine in some cases due to the small map sizes.  

In addition, several regions had no cores identified
but displayed hints of structures with peaks below the object-identification 
theshold of 5$\sigma_{C}$.  In order to put some constraints / 
upper limits on the submillimetre properties of potential cores in these
regions, we ran clumpfind to a lower identification threshold 
($\sigma_{C}$ = 0.01~Jy~beam$^{-1}$ per pixel) in these regions only.  
It should 
be noted that properties derived for these objects are not as reliable 
as the originally identified cores.  We term these objects as having 
`less secure fits' throughout the paper and do not include these in our 
quantitative analysis.  Table~\ref{scuba_appendix} below denotes the
properties of the SCUBA cores identified.

Finally, as discussed in Appendix~C, SCUBA observations in the region of NGC1333
appear to have a shift of 6\arcsec\ in RA relative to data at other
wavelengths, apparently due to an unusually large pointing error
at the JCMT.  Here, we apply a global shift of 6\arcsec\ to the
NGC1333 observations to compensate for this.

\subsection{Comparison to Previous Results}
While Clumpfind does a reasonable job of identifying structures in two 
dimensional maps where the filling factor is low, the cores identified in
the 3\arcsec\ sampled maps used here are different from the set we 
identified previously in the 6\arcsec\ sampled map of \citetalias{Kirk06}, 
even though the same data {\hk are}
used and the reduction procedure is almost identical, albeit with
a different smoothing scale.  This is
because Clumpfind relies on contours for determining object edges, so that
slight variations in flux per pixel can change the size of core boundaries, 
which then affects the measured size and total flux, although the peak
flux would be unchanged.  In clustered regions, the slight variation in 
flux per pixel can
also change where or when cores are either separated from or merged with 
close neighbours.  Clumpfind identifies distinct clumps where two regions are
isolated at a given search contour (every 2$\sigma_{C}$).
For example, a peak at 6.9$\sigma_{C}$ surrounded by pixels at 6.1$\sigma_{C}$ 
and near a peak at 11$\sigma_{C}$ would be identified as a single object 
(at the 7 and 5$\sigma_{C}$ contours, all of the flux is connected).  If
the 6.9$\sigma_{C}$ peak were instead a peak at 7.1$\sigma_{C}$, 
it would be identified as a separate object from the 11$\sigma_{C}$ peak 
(the flux is in two unconnected regions at the 7$\sigma_{C}$ 
contour).  Thus individual core properties are difficult to compare
between maps reduced under even slightly different schemes or
resolutions.  We do not show a comparison
of the cores identified here with those identified in the 6\arcsec\ sampled
map of \citetalias{Kirk06}, but note that given variations in
core boundaries and potential merging of cores, the list of cores we 
identify in the 3\arcsec\ sampled map spans the cores identified in the 
6\arcsec\ sampled map and also includes an additional four cores
(\#4, 28, 72, 73).

\subsection{Core Properties}
The properties of the cores identified in the 3\arcsec\ sampled map are
shown in Table~\ref{scuba_appendix}.  The core radius, peak flux, and total flux
are found with Clumpfind.  We also calculate the concentration of
each core -- previous work has shown this to be an indicator
of the evolutionary state of the core, with higher concentrations
corresponding to later stages of evolution
\citep{Walawend05,Walawend06,Johnstone06}.
Following \citet{Johnstone01}, the concentration can be calculated
from observational quantities as: 
\begin{equation}
C = 1 - \frac{1.13 B^{2}S_{850}}{(\pi R^{2}_{obs})f_{0}}
\end{equation}
where B is the beamsize, $S_{850}$ the total flux, $R_{obs}$ the
radius, and $f_{0}$ the peak flux.  As in \citetalias{Kirk06}, we can
also model the cores as Bonnor Ebert (BE) spheres -- spherically symmetric
isothermal objects where thermal pressure balances gravity and an 
external pressure.  The best fit BE sphere model properties are also
included in Table~\ref{scuba_appendix}.

\section{TWO-COMPONENT \nh CORES}

In this section we discuss the \nh spectra which we found required two
velocity components for a good model fit.
We treated these two velocity components as originating from independent
objects along the line of sight, rather that a single core whose spectrum
shows self-absorption.  The optimal method for distinguishing between
self-absorption and two distinct cores would be to observe the region
with an optically thinner tracer.  Since this is not available to us, we
instead examine the data we do have for these cores (e.g. spectra at
cross positions, SCUBA observations, and other \nh survey data) and
discuss how well they support our interpretation. 

Most of the cores for which we fit to two components lie in regions where
complex motions are seen on smaller scales, such as NGC1333, where it
is not unsurprising to find two cores along the line of sight.  
In instances where we found a common velocity component between two
close pointings, we excluded this common velocity component from
the second source in our subsequent analysis in order
to avoid counting the same core twice.

\begin{figure}
\plotone{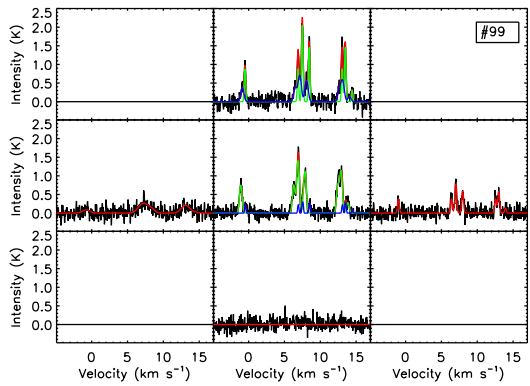}
   \caption{Spectrum for source \#99 in NGC1333.  The vertical axis is in
	units of T$_{A}^{*}$.  Red indicates the summed fit while blue
	and green indicate the two different components fit (where applicable).}
   \label{fig_ps_s36}
\end{figure}

In NGC1333, we found six cores with two
velocity components -- \#99, 103, 106, 107, 111, and 118. 
All of these except \#99 fall within the survey region
of \citet{Walsh06} which in every instance identifies a distinct
object at each velocity which we fit to our data.  Source \#99, which falls
outside of the \citet{Walsh06} survey region, was observed at cross positions 
(see Figure~\ref{fig_ps_s36}).  The two components show 
varying relative intensities across the
five positions mapped, supporting the hypothesis of two distinct cores.

\begin{figure}
\plotone{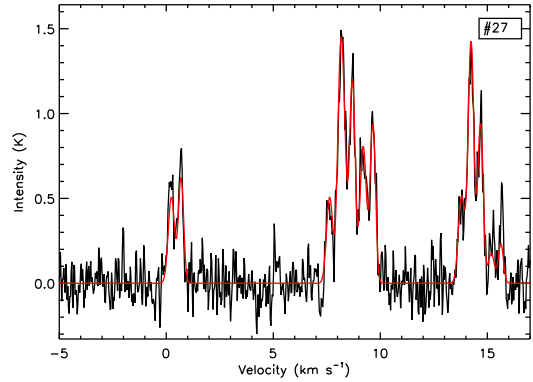}
  \caption{Spectrum for source \#27 in IC348.  The same plotting convention
	is used as in Figure~\ref{fig_ps_s36}.}
\label{IC348}
\end{figure}

In IC348, we fit two cores with two velocity components, \#22 and 27.  
While there is no high resolution \nh map of IC348, \citet{Tafalla06}
provide a 50\arcsec\ resolution map.  Our core \#22 corresponds to
\citet{Tafalla06}'s source C in IC348-SW1 which they show appears to
be related to outflow structure seen in CO.  \citet{Tafalla06}
identify two velocity components in \nh which correspond to the velocities
we found.  Our core \#27 corresponds to source A in IC348-SW1
which \citet{Tafalla06} associate with a single broad ($\sim$1.6~km/s)
velocity component.  Our observations (Figure~\ref{IC348}) clearly 
show two distinct
velocity components, perhaps resolvable due to the smaller beamsize of
our observations.

\begin{figure}
\plotone{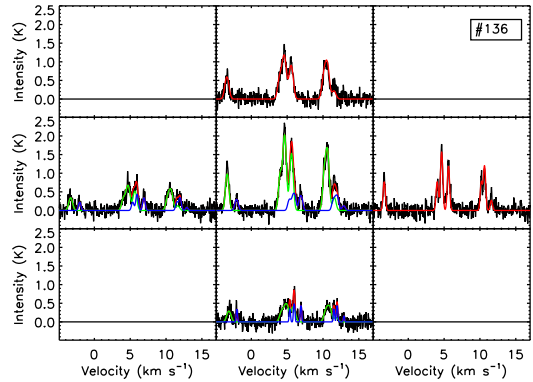}
  \caption{Spectrum for source \#136 in L1544.  The same plotting convention
	is used as in Figure~\ref{fig_ps_s36}.}
\label{fig_ps_s41}
\end{figure}

In L1544, one core (\#136) was fit with two velocity components.
Cross positions were also observed, and where the S/N is high enough, 
the 
two components are clearly quite separate (see Figure~\ref{fig_ps_s41}).

North of B1, one core (\#76) was fit with two velocity components at
one cross position only.  This observation has low S/N and is designated as 
being of poor quality, and therefore was not used in any of the analysis.

\begin{figure}
\plotone{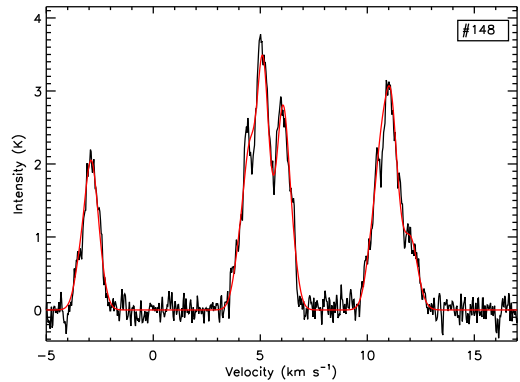}
  \caption{Spectrum of source \#148, the two velocity component core in L1448.
	The same plotting convention is used as in Figure~\ref{fig_ps_s36}.}
\label{L1448}
\end{figure}

The core with the poorest two velocity component fit is \#148 in L1448.
In this instance, the model does not well fit the data in several places
(see Figure~\ref{L1448}), making the interpretation of two distinct cores 
uncertain.  The SCUBA map suggests the core is isolated -- the nearest
SCUBA core is roughly an arcminute away.  Without
additional observations, no firm interpretation
can be made.  

\section{EFFECT OF UTILIZING SINGLE POINTINGS}
For the analysis in the paper, we utilized only the data from
the central pointing on each target in order to have a consistent
dataset.  Since we have four-point cross data around some of our
targets, we can ask two additional questions -- how accurate were
the target centre positions that we used and does our utilization
of only the central point bias or change any of our results.  Here we
address these questions in turn.

\subsection{Accuracy of Dense Core Pointings}
We first note one occurence that led to small offsets between some
positions in the SCUBA catalog we used for determining our target 
positions and the catalog of Appendix~A.  
After the publication of \citetalias{Kirk06}, we discovered an offset 
of $\sim$6\arcsec\ in RA 
in the region of NGC1333 in the SCUBA data compared with data 
at other wavelengths.  This is intrinsic to the data (not caused by
an analysis error) and is apparently caused by an unusually large 
pointing error with SCUBA.  This offset is further discussed in 
\citet{DiFrancesco07}.  This offset was not known at the time of
our IRAM observations, hence was not accounted for when the 
SCUBA target list was created.
This 6\arcsec\ offset is much smaller 
than the IRAM beamsize (25\arcsec) and hence should not have a large 
effect on the results, but is noted here for completeness.

We expect the \nh cores to have extents of order one or two 
IRAM beams (25\arcsec) in \nhns, the typical size of the
SCUBA cores.  SCUBA is sensitive to a similar range of densities, 
$\sim 10^{4}-10^{6}$~cm$^{-3}$ using the approximation
in \S7.  If we chose our target positions well, we expect the central 
position to show more signal than offset positions, although the cores
should extend past the central pointing.

\begin{figure}
\plotone{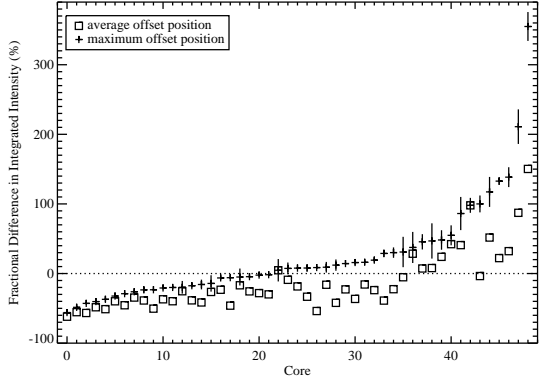}
   \caption{Fractional difference in integrated intensity between 
	centre and offset positions in dense \nh cores.  The pluses denote the
	maximum difference with the vertical lines indicating the error
	in this measurement.  The squares denote the average difference
	for all four offset positions versus the centre.
	The cores are ordered according to the maximum difference.}
   \label{fig_cross_area}
\end{figure}

Figure~\ref{fig_cross_area} shows the fractional
difference between the integrated intensity at the central position
and largest value at an offset position (pluses) as well as the
difference with the average value of all offset positions (squares).
The vertical lines indicate the error in the difference measure for
the maximum difference.  Nearly half of the cores plotted
have highest integrated intensities at the central pointing,
while another third have integrated intensities which are slightly
larger at a single cross position.  Only about one fifth of the cores 
show significantly larger integrated intensities at a single offset position.
An additional thirteen positions where offset observations were made are not
plotted due to a lack of a good central detection.  Of these, ten also
had no detections at any offset positions.

It should be noted that we utilized two different set of criteria to
determine which targets to map crosses around.
Approximately half (thirty of sixty-two) of the cross targets
had strong central detections -- we observed at offsets to search for  
extended structure.  The remaining half of the targets
were chosen based on a very
weak or non-existent central detection in order to search for a 
stronger signal nearby (2MASS or Palomar plate - selected
targets).  The first of these cross-map criteria biases
towards targets where the centre was well chosen, while the second 
biases towards poorly chosen centres.
The former set of cross maps were mostly based on SCUBA cores
where we had a high detection probability and it was easiest to
determine precise core centres.  Nearly all of the SCUBA-selected 
cores with crosses are consistent with having 
their strongest integrated intensity at the central position.
The latter-chosen cross targets account for the largest integrated
intensity differences at
cross positions as well as the targets with insufficient signal to noise to
obtain a fit and thus be included in the plot.
Not surprisingly, our position selection was less accurate for these 
latter targets -- many of the detections are not consistent with
the central position having the highest integrated intensity.  
Since these latter crossed observations
are biased to the worst-determined centre positions, we expect
that our overall determination of centres was more successful.

We can therefore conclude that the pointings were quite accurate for the
SCUBA-selected targets and reasonably accurate for the targets selected 
using the
other methods, especially given the difficulties
in determining precise centres in the latter case.  As we shall see in the
following section, the question of accuracy of the centre positions does
not play an important role in the kinematical results presented earlier --
these change very little between centre and cross positions.

\subsection{Impact on Results}
Now we examine whether our utilization of a single pointing of each dense
\nh core introduces significant bias or error in the core properties
measured.

\subsubsection{Line Widths}

We calculated the difference in \nh linewidth between each centre position
and the surrounding cross and found this difference to be very small.
Most (93\%) of the \nh cores have mean differences less than one or two spectral
channels (0.05~km/s in FWHM units), i.e., consistent with measuring
the same value.  The mean absolute difference of the entire sample was 
0.018~km/s and none had differences close to
the sound speed (FWHM of 0.15~km/s for \nhns) and the mean
difference was approximately zero.  Therefore, we expect 
that our analysis utilizing only central pointings does not introduce 
significant error or bias to our measurement of core linewidth.

\subsubsection{Centroid Velocities}
\begin{figure}
\plotone{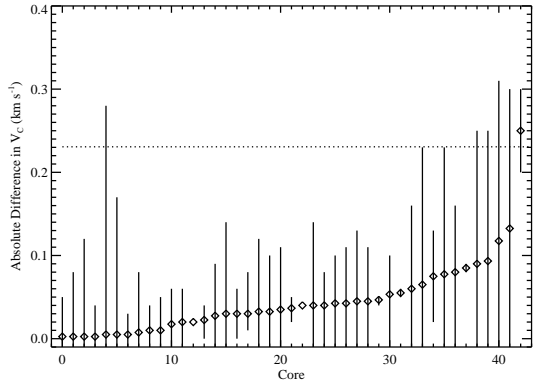}
  \caption{Difference in centroid velocity between the cross and central
	positions, with \nh cores ordered in increasing average difference.  
	The diamonds indicate the average cross velocity difference
	while the vertical lines indicate the range in centroid velocity 
	difference at all cross positions.  Cores with two velocity
	components fit have not been included.  The dotted line
	indicates the sound speed.
	}
  \label{fig_core_core_cross}
\end{figure}

The difference in centroid velocity for the \nh cores at centre and
cross positions is also small, as shown in Figure~\ref{fig_core_core_cross}.
In Figure~\ref{fig_core_core_cross}, the vertical lines indicate the
range in differences over the four cross positions.
Note that the six sources with two velocity components
fit at either centre or cross position have not been included in this plot.
All but one of the \nh cores have a difference between the central and average
cross velocity of less than the thermal velocity dispersion; 
the mean difference is approximately zero and the standard deviation
is 0.04~km~s$^{-1}$.  The \nh cores which have
the largest centroid velocity differences have less secure fits where 
there is greater uncertainty in the centroid velocity determined.
On the other hand, in the majority of cases
the difference in velocities between cross and central positions
are larger than the the fitting errors of those velocities,
indicating that there is real variation in the centroid velocity
between locations.  This variation is much smaller
than the core-to-core variations analyzed in the \S6, however,
where extinction regions require support against gravity with
core-to-core motions many times larger than the sound speed.
We thus find that little error or bias is introduced into our results
by using data from a central pointing only.


\begin{deluxetable}{cccccccccc}
\tabletypesize{\scriptsize}
\tablewidth{0pt}
\tablecolumns{10}
\tablecaption{Target Properties \label{targets}}
\tablehead{
\colhead{Number} &
\colhead{RA} &
\colhead{Dec} &
\colhead{Sel. \tablenotemark{a}} &
\colhead{Cross? \tablenotemark{b}} &
\colhead{SCUBA \tablenotemark{c}} &
\colhead{SCUBA \tablenotemark{d}} &
\colhead{Spitzer \tablenotemark{e}} &
\colhead{Ext.\ \tablenotemark{f}} &
\colhead{Mean Ext.\ \tablenotemark{g}}
\\
\colhead{} &
\colhead{(J2000.0)} &
\colhead{(J2000.0)} &
\colhead{Method} &
\colhead{} &
\colhead{core} &
\colhead{flux (Jy)} &
\colhead{protostar} &
\colhead{Region} &
\colhead{(A$_{V}$)} 
}
\startdata
   1 & 3:48:31.1 &32:38:21.0 &   P &  no &.. &    0 &.. &... &  1.9 \\
   2 & 3:48:29.1 &32:54:33.0 &   P &  no &.. &    0 &.. &  1 &  4.2 \\
   3 & 3:47:42.3 &32:52:28.1 &   E & yes &.. & 0.88 &.. &  1 &  8.1 \\
   4 & 3:47:41.8 &32:51:40.3 &   S &  no & 1 & 1.74 &49 &  1 &  8.0 \\
   5 & 3:47:40.4 &32:54:39.0 &   P & yes &.. & 0.28 &.. &  1 &  8.2 \\
   6 & 3:47:39.0 &32:52:11.2 &   S &  no & 3 & 1.46 &.. &  1 &  8.1 \\
   7 & 3:47:37.0 &32:52:03.0 &   P & yes &.. & 0.28 &.. &  1 &  7.8 \\
   8 & 3:47:36.1 &32:50:11.0 &   P & yes &.. &    0 &.. &  1 &  7.1 \\
\enddata
\tablenotetext{a}{Target selection was based on SCUBA cores (S), potential SCUBA cores (SA), Palomar plate candidates (P), and 2MASS extinction structures (E).}
\tablenotetext{b}{Were offset positions observed?}
\tablenotetext{c}{SCUBA core reference number (see Appendix)} 
\tablenotetext{d}{Total SCUBA flux within the IRAM beam \citep[derived from the 6\arcsec\ map of ][]{Kirk06}.}
\tablenotetext{e}{Spitzer reference number \citep[from][]{Jorgensen07} }
\tablenotetext{f}{Extinction region number \citep[from][]{Kirk06}}
\tablenotetext{g}{Mean extinction within the IRAM beam.}
\end{deluxetable}

\begin{deluxetable}{ccccc}
\tabletypesize{\small}
\tablewidth{0pt}
\tablecolumns{5}
\tablecaption{Detection rates for the target selection methods.\label{table_detect}}
\tablehead{
\colhead{Tracer} &
\colhead{SCUBA} &
\colhead{Palomar} &
\colhead{2MASS} &
\colhead{Total\tablenotemark{a}}
}
\startdata
\nh(1-0) & 84\% (75/89)& 42\% (18/43)& 14\%  (3/22)& 62\%  (96/154) \\
\co(2-1) & 98\% (87/89)& 93\% (40/43)& 95\% (21/22)& 96\%  (148/154)\\
\enddata
\tablenotetext{a}{Several targets in the Palomar- and 2MASS- selected 
lists lie close to a SCUBA-selected target and appear to sample the
same object (one and two targets respectively).  We only count these 
targets once for both these statistics and the subsequent analysis.}
\end{deluxetable}

\begin{deluxetable}{cccccccccccc}
\tabletypesize{\scriptsize}
\tablewidth{0pt}
\tablecolumns{12}
\tablecaption{Parameters of Spectral Fitting for \nh \label{table_fits_nh} }
\tablehead{
\colhead{\#\tablenotemark{a}} &
\colhead{RA\tablenotemark{b}} & 
\colhead{dec\tablenotemark{b}} & 
\colhead{$V_{C}$\tablenotemark{c,d}} &
\colhead{$\Delta V$\tablenotemark{c}} &
\colhead{error\tablenotemark{c}} &
\colhead{optical\tablenotemark{c}} & 
\colhead{error\tablenotemark{c}} & 
\colhead{rms\tablenotemark{c}} &
\colhead{rms\tablenotemark{c}} &
\colhead{$\int$ I \tablenotemark{e}} & 
\colhead{secure\tablenotemark{f}}
\\
\colhead{} &
\colhead{offset} & 
\colhead{offset} & 
\colhead{(km~s$^{-1}$)} & 
\colhead{(km~s$^{-1}$)} &
\colhead{(m~s$^{-1}$)} &
\colhead{depth} &
\colhead{} &
\colhead{base(K)} &
\colhead{fit(K)} &
\colhead{(K~km~s$^{-1}$)} &
\colhead{fit ?} 
}
\startdata
   1 &  0 &  0 &   ... &   ... &   ... &   ... &   ... &  0.15 &   ... &   ... & ... \\
   2 &  0 &  0 &   ... &   ... &   ... &   ... &   ... &  0.15 &   ... &   ... & ... \\
   3 &  0 &  0 & 10.22 &  0.25 &     5 &  6.53 &  0.64 &  0.08 &  0.10 &  2.38 & yes \\
   3 &  0 & -1 & 10.26 &  0.40 &     6 &  5.41 &  0.48 &  0.12 &  0.15 &  5.54 & yes \\
   3 &  0 &  1 & 10.21 &  0.28 &    12 &  1.75 &  1.19 &  0.12 &  0.11 &  1.40 & yes \\
   3 & -1 &  0 & 10.23 &  0.36 &     7 &  5.29 &  0.60 &  0.12 &  0.17 &  3.71 & yes \\
   3 &  1 &  0 & 10.22 &  0.28 &     8 &  0.88 &  0.50 &  0.12 &  0.11 &  0.98 & yes \\
   4 &  0 &  0 & 10.24 &  0.45 &     5 &  6.30 &  0.33 &  0.10 &  0.18 &  7.39 & yes \\
\enddata
\tablenotetext{a}{See Table~\ref{targets} for the locations of the targets.}
\tablenotetext{b}{measured in units of the beam (25\arcsec)}
\tablenotetext{c}{Centroid velocity, velocity dispersion (FWHM), optical depth, baseline rms, model rms, and the corresponding errors where appropriate determined using CLASS}
\tablenotetext{d}{The error was similar in all cases with a mean and standard deviation of 8$\pm$8~m/s.}
\tablenotetext{e}{Integrated intensity computed using the tdv function in CLASS.}
\tablenotetext{f}{Was the fit judged to be good?}
\end{deluxetable}

\begin{deluxetable}{ccccccccccc}
\tabletypesize{\scriptsize}
\tablewidth{0pt}
\tablecolumns{11}
\tablecaption{Parameters of Spectral Fitting for \co \label{table_fits_co} }
\tablehead{
\colhead{\#\tablenotemark{a}} &
\colhead{RA\tablenotemark{b}} & 
\colhead{dec\tablenotemark{b}} & 
\colhead{$V_{C}$\tablenotemark{c,d}} &
\colhead{$\Delta V$\tablenotemark{c}} &
\colhead{error\tablenotemark{c}} &
\colhead{T$_{A,peak}^{*}$\tablenotemark{c}} & 
\colhead{$\int$ I \tablenotemark{c}} & 
\colhead{rms\tablenotemark{c}} &
\colhead{rms\tablenotemark{c}} &
\colhead{secure\tablenotemark{e}}
\\
\colhead{} &
\colhead{offset} & 
\colhead{offset} & 
\colhead{(km~s$^{-1}$)} & 
\colhead{(km~s$^{-1}$)} &
\colhead{(m~s$^{-1}$)} &
\colhead{(K)} &
\colhead{(K~km~s$^{-1}$)} &
\colhead{base(K)} &
\colhead{fit(K)} &
\colhead{fit ?} 
}
\startdata
   1 &  0 &  0 &   ... &   ... &   ... &   ... &   ... &  0.47 &   ... &   ...\\
   2 &  0 &  0 & 10.42 &  0.46 &    51 &  2.32 &  1.14 &  0.42 &  0.36 & yes \\
   2 &  0 &  0 &  9.61 &  0.72 &   103 &  1.50 &  1.15 &  0.42 &  0.36 & yes \\
   3 &  0 &  0 & 10.20 &  0.53 &    11 &  3.92 &  2.21 &  0.18 &  0.35 & yes \\
   3 &  0 & -1 & 10.13 &  0.84 &    19 &  3.00 &  2.68 &  0.23 &  0.46 & yes \\
   3 &  0 &  1 & 10.16 &  0.50 &    10 &  4.68 &  2.49 &  0.25 &  0.27 & yes \\
   3 &  0 &  1 &  9.21 &  0.40 &    38 &  1.54 &  0.65 &  0.25 &  0.27 & yes \\
   3 & -1 &  0 & 10.21 &  0.58 &    12 &  4.86 &  3.02 &  0.27 &  0.38 & yes \\
\enddata
\tablenotetext{a}{See Table~\ref{targets} for the locations of the targets.}
\tablenotetext{b}{measured in units of the beam (25\arcsec)}
\tablenotetext{c}{Centroid velocity, velocity dispersion (FWHM), peak intensity, integrated intensity, baseline rms, model rms, and the corresponding errors where appropriate determined using CLASS}
\tablenotetext{d}{The error was similar in all cases with a mean and standard deviation of 16$\pm$15~m/s.}
\tablenotetext{e}{Was the fit judged to be good?}
\end{deluxetable}

\begin{deluxetable}{ccccccccccc}
\tabletypesize{\scriptsize}
\tablewidth{0pt}
\tablecolumns{11}
\tablecaption{Properties of Extinction Regions.\label{extinct_regions}}
\tablehead{
\colhead{Number \tablenotemark{a}} &
\colhead{RA\tablenotemark{a}} &
\colhead{Dec\tablenotemark{a}} &
\colhead{Mass\tablenotemark{a}} &
\colhead{Radius\tablenotemark{a}} &
\colhead{$\sigma V_{grav}$\tablenotemark{b}} &
\colhead{N$_{N}$\tablenotemark{c}} &
\colhead{$\sigma V_{N}$\tablenotemark{d}} &
\colhead{N$_{C}$\tablenotemark{e}} &
\colhead{$\sigma V_{C}$\tablenotemark{f}} &
\colhead{Notes \tablenotemark{g}}
\\
\colhead{} &
\colhead{(J2000.0)} &
\colhead{(J2000.0)} &
\colhead{(M$_{\odot}$)} &
\colhead{(\arcsec)} &
\colhead{(km~s$^{-1}$)} &
\colhead{ } &
\colhead{(km~s$^{-1}$)} &
\colhead{ } &
\colhead{(km~s$^{-1}$)} &
\colhead{ }
}
\startdata
  1& 3:47:45.3& 32:52:43.4&  859.6&  776& 0.89& 7& 0.39&  9& 0.46&B5\\
  2& 3:43:57.1& 31:59:28.7& 1938.9& 1119& 1.11&11& 0.55& 25& 0.51&IC348\\
  3& 3:39:27.4& 31:21:08.6&  780.6&  737& 0.87& 2& 0.20& 11& 0.41&SW of IC348\\
  4& 3:36:28.9& 31:11:13.1&  560.5&  670& 0.77& 1& 0   &  5& 0.42&E of B1\\
  5& 3:32:35.6& 30:58:27.7&  441.1&  579& 0.73& 9& 0.28& 19& 0.43&B1\\
  6& 3:30:28.7& 30:26:30.2&  257.6&  454& 0.63& 3& 0.09&  4& 0.30&SW of B1\\
  7& 3:28:56.0& 31:22:36.4&  973.3&  889& 0.88& 8& 0.54& 12& 0.78&NGC1333\\
  8& 3:28:53.3& 30:44:00.5&  246.2&  453& 0.62& 0& 0   &  2& 0.30&S of NGC1333\\
\enddata
\tablenotetext{a}{Extinction region information from \citetalias{Kirk06}}
\tablenotetext{b}{Velocity dispersion required to support the region}
\tablenotetext{c}{Number of starless cores detected in \nh within the 
	extinction region} 
\tablenotetext{d}{Velocity dispersion of the \nh starless cores in the region 
	(uncorrected for the thermal motion of the mean gas)}
\tablenotetext{e}{Number of non-protostars detected in \co within the
	extinction region}
\tablenotetext{f}{Velocity dispersion of the non-protostellar \co detections
	in the extinction region (uncorrected for the thermal motion of the 
	mean gas)}
\tablenotetext{g}{Descriptive location of extinction region}
\end{deluxetable}

\clearpage
\begin{deluxetable}{ccccccccccccc}
\rotate
\tabletypesize{\scriptsize}
\tablewidth{0pt}
\tablecolumns{13}
\tablecaption{Properties of the submillimetre cores identified in the Perseus SCUBA map with 3\arcsec\ pixels.\label{scuba_appendix}}
\tablehead{
\colhead{Name \tablenotemark{a}} &
\colhead{\# \tablenotemark{b}} &
\colhead{RA \tablenotemark{c}} &
\colhead{Dec \tablenotemark{c}} &
\colhead{f$_{0}$ \tablenotemark{d}} &
\colhead{S$_{850}$ \tablenotemark{d}} &
\colhead{R$_{eff}$ \tablenotemark{d}} &
\colhead{Conc \tablenotemark{e}} &
\colhead{T$_{BE}$ \tablenotemark{e}} &
\colhead{log $n_{cent}$\tablenotemark{e}} &
\colhead{P$_{ext}$/k \tablenotemark{e}} &
\colhead{Secure \tablenotemark{f}} &
\colhead{Assoc \tablenotemark{g}} 
\\
\colhead{(SMM J)} &
\colhead{} &
\colhead{(J2000.0)} &
\colhead{(J2000.0)} &
\colhead{(Jy/bm)} &
\colhead{(Jy)} &
\colhead{(\arcsec)} & 
\colhead{} &
\colhead{(K)} &
\colhead{(cm$^{-3}$)} &
\colhead{(cm$^{3}$~K$^{-1}$)} &
\colhead{Fit?} &
\colhead{Protostar}
}
\startdata
034769+32517 &  1 & 3:47:41.5 &32:51:46.5 &  0.4 & 1.07 & 22.9 &0.59 & 11. & 5.68 & 6.00 & yes & 49 \\
034765+32530 &  2 & 3:47:39.2 &32:53:01.5 &  0.2 & 0.26 & 13.7 &0.20 & 15. & 5.14 & 6.20 & yes & -- \\
034765+32522 &  3 & 3:47:39.2 &32:52:16.5 &  0.4 & 0.74 & 18.9 &0.49 & 12. & 5.54 & 6.20 & yes & -- \\
034481+32005 &  4 & 3:44:48.8 &32:00:30.5 &  0.2 & 0.40 & 17.2 &0.29 & 16. & 4.99 & 6.10 & yes & -- \\
034473+32015 &  5 & 3:44:43.9 &32:01:33.7 &  0.6 & 1.08 & 20.8 &0.62 & 12. & 5.88 & 6.10 & yes & 48 \\
034461+31588 &  6 & 3:44:36.8 &31:58:48.8 &  0.2 & 0.49 & 18.1 &0.39 & 14. & 5.11 & 6.10 & yes & -- \\
034438+32100 &  7 & 3:44:23.1 &32:10:01.3 &  0.1 & 0.30 & 18.6 &0.32 & 14. & 4.85 & 5.90 &  no & -- \\
034410+32022 &  8 & 3:44:06.1 &32:02:16.4 &  0.2 & 1.30 & 28.0 &0.38 & 18. & 4.79 & 5.90 & yes & -- \\
\enddata
\tablenotetext{a}{Name formed from J2000 positions (hhmm.mmddddmm.m)}
\tablenotetext{b}{Reference number used in the text of this paper}
\tablenotetext{c}{Position of peak flux within core (accurate to 3\arcsec).}
\tablenotetext{d}{Peak flux, total flux, and radius derived from {\it clfind} \citep{Williams94}.  The beamsize is 15.7\arcsec\ for the peak flux.}
\tablenotetext{e}{Concentration, temperature, central number density, and external pressure derived from Bonnor-Ebert modelling (see text).}
\tablenotetext{f}{See text for description of which submillimetre core identificationsare considered to be secure.}
\tablenotetext{g}{The reference number of any associated Spitzer protostar from the catalog of \citet{Jorgensen07}.  Note that if more than one protostar is considered to be associated (within 15\arcsec), only the closest one is listed.}
\end{deluxetable}

\clearpage

\clearpage

\clearpage


\end{document}